\documentclass{article}

\usepackage{arxiv}

\usepackage[utf8]{inputenc} 
\usepackage[T1]{fontenc}    
\usepackage{url}            
\usepackage{booktabs}       
\usepackage{amsfonts}       
\usepackage{nicefrac}       
\usepackage{microtype}      
\usepackage{lipsum}		
\usepackage{graphicx}
\usepackage{doi}

\usepackage[utf8]{inputenc} 
\usepackage[utf8]{inputenc}
\usepackage{graphicx,epsfig}
\usepackage{url}
\usepackage{xcolor}
\usepackage{color}
\usepackage{multirow}
\usepackage{tablefootnote}
\usepackage{xspace}
\usepackage{hyperref}
\usepackage{subfig}
\usepackage{tabularx,ragged2e,booktabs,caption}
\usepackage{graphics}

\title{The structure of segregation in co-authorship networks and its impact on scientific production}

\author{\href{https://orcid.org/0000-0003-2409-3064}{\includegraphics[scale=0.06]{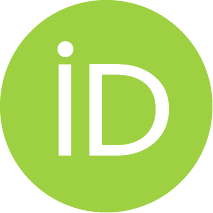}\hspace{1mm}Ana Maria Jaramillo} \\
    BioComplex Laboratory\\
	Department of Computer Science\\
	University of Exeter, UK\\
	\texttt{ajaramillo@biocomplexlab.org} \\
	\And
	\href{https://orcid.org/0000-0002-5927-3367}{\includegraphics[scale=0.06]{orcid.pdf}\hspace{1mm} Hywel T.P. Williams} \\
	SEDA Lab\\
	Department of Computer Science\\
	University of Exeter, UK \\
	\AND
	\href{https://orcid.org/0000-0002-5559-3064}{\includegraphics[scale=0.06]{orcid.pdf}\hspace{1mm}Nicola Perra} \\
	School of Mathematical Sciences\\
	Queen Mary University of London, UK\\
	\AND
	\href{https://orcid.org/0000-0002-6479-6429}{\includegraphics[scale=0.06]{orcid.pdf}\hspace{1mm}Ronaldo Menezes} \\
	BioComplex Laboratory\\
	Department of Computer Science\\
	University of Exeter, UK \\
	\texttt{r.menezes@exeter.ac.uk}
}

\renewcommand{\shorttitle}{The structure of segregation in co-authorship networks and its impact on scientific production}
\newcommand{\sectionname}{Section}

\hypersetup{
pdftitle={The structure of segregation in co-authorship networks and its impact on scientific production},
pdfsubject={CS},
pdfauthor={Ana Maria Jaramillo, Hywel T.P. Williams, Nicola Perra, Ronaldo Menezes},
pdfkeywords={science of science, coreness, collaboration networks, segregated communities}
}

\begin{document}
\maketitle

\begin{abstract}
Co-authorship networks, where nodes represent authors and edges represent co-authorship relations, are key to understanding the production and diffusion of knowledge in academia. Social constructs, biases (implicit and explicit), and constraints (e.g. spatial, temporal) affect who works with whom and cause co-authorship networks to organise into tight communities with different levels of segregation. We aim to look at aspects of the co-authorship network structure that lead to segregation and its impact on scientific production. We measure segregation using the Spectral Segregation Index (SSI) and find 4 ordered segregation categories: completely segregated, highly segregated, moderately segregated and non-segregated communities. We direct our attention to the non-segregated and highly segregated communities, quantifying and comparing their structural topologies and k-core positions. When considering communities of both categories (controlling for size), our results show no differences in density and clustering but substantial variability in core position. Larger non-segregated communities are more likely to occupy cores near the network nucleus, while the highly segregated ones tend to be closer to the network periphery. Finally, we analyse differences in citations gained by researchers within communities showing different segregation categories. Researchers in highly segregated communities get more citations from their community members in middle cores and gain more citations per publication in middle/periphery cores. Those in non-segregated communities get more citations per publication in the nucleus. To our knowledge, this work is the first to characterise community segregation in co-authorship networks and investigate the relationship between community segregation and author citations. Our results help study highly segregated communities of scientific co-authors and can pave the way for intervention strategies to improve the growth and dissemination of scientific knowledge.
\end{abstract}


\section{Introduction}

The social structures behind scientific production may have profound effects on the growth and dissemination of knowledge, the well-being of our societies, and the evolution of academic research~\cite{Fortunato2018}. Many studies have shown how socially influenced behaviours impact different aspects of the scientific enterprise. Examples include the selection of co-authors, citation rates, and peer review processes, which are biased by author attributes such as prestige~\cite{Lynn2014}, gender~\cite{Sugimoto2013}, and country of affiliation~\cite{Smith2014,Opthof2002}.
 
Co-authorship networks, where nodes represent researchers and links represent co-authorship relations between them, have been shown as key to the understanding and mapping of scientific production~\cite{Zeng2017, Pan2012, Pan2018}. Particular attention has been devoted to their structural properties. These networks are organised in communities formed by groups of highly collaborative researchers with relatively low external interactions~\cite{Newman2006}. Looking at the evolution of these networks in time, one might see these communities going from being disconnected components to joining the giant component, as the co-authorship network coalesces. When comparing the proportion of nodes in the giant component relative to the total number of nodes, critical transition points represent the constitution of new disciplines and the growth of science~\cite{Bettencourt2009}.

As in most activities driven by human interactions, the biases mentioned above influence the processes of community formation and their connection/disconnection with other parts of the network. On one side, the previous literature has shown how the lack of exposure to individuals outside their circle can create segregated groups~\cite{SUNSTEIN2018}. In different contexts of scientific production, such as discussions on social media, this ``structural segregation''~\cite{Kim2019} can increase polarization~\cite{Sasahara2021,Perra2019} and reinforce similar opinions ~\cite{DelVicario2016}. High segregation levels---found in social networks with very fragmented groups---hamper the development of social capital and the emergence of cooperative behaviour, to the detriment of innovation, social learning, and problem solving~\cite{Henry2011}. In particular, computer scientists immersed in gender-segregated groups (low female-male connectivity) have disadvantaged positions in accessing information ~\cite{Jalali2020}. On the other side, researchers grouped into segregated communities could increase the exploitation of innovative ideas with in-depth work. For example, groups of researchers organised in efficient structures, characterised for being more interconnected and less clustered, proved to outperform others in solving complex problems~\cite{Mason2012}, and researchers from evolutionary medicine produce better and longer-lasting ideas when located on the network's periphery~\cite{Painter2021}. There is tension between consolidating and diversifying collaborations, as both might affect the growth of scientific knowledge and research impact. Our understanding of when and how collaborations across communities can help expand research methods and questions~\cite{Nielsen2018}, as well as promote the spreading of scientific results~\cite{Smith2014, Sonnenwald2007}, is still limited.

In this context, we tackle 3 specific research questions: {\em (i)} How to identify highly segregated communities in co-authorship networks? {\em (ii)} Are there differences in the topological structure and core position of communities with different segregation levels? {\em (iii)} Does the segregation level affect success in science as measured by citations?

To answer these questions, we study co-authorship networks using a dataset of publications in Computer Science. We assume that communities of researchers with very high internal connectivity versus low external connectivity can be considered highly segregated. We use 4 ordered segregation categories and show a relationship between community size, segregation category, and core position, whereby non-segregated communities tend to be positioned near the network's nucleus. We also find that highly segregated researchers gain more citations when positioned in the middle or periphery cores of the network. In comparison, non-segregated researchers gain more citations in cores near the nucleus. Also, highly segregated researchers gain a higher proportion of their citations from their own communities in middle cores, while non-segregated researchers do so in the nucleus.
 
The paper is organised as follows: \sectionname~\ref{sec:data} describes the dataset and network properties used in this study. \sectionname~\ref{sec:Community_partition} details the procedure and characterisation of the community partition. \sectionname~\ref{sec:Communities_segregation} defines the structural segregation metric used in this study and how communities are categorised as completely segregated, highly segregated, moderately segregated and non-segregated. Our analyses focus on understanding non-segregated and highly segregated communities. \sectionname~\ref{sec:Communities_topology} shows 4 metrics related to the topology and core position of these communities, and we compare them using distributions and Z-Scores. In \sectionname~\ref{sec:Citations_researchers}, we compare the number of citations per publication, and the proportion of citations received by members of the same community, to analyse the implications for researchers in communities with different segregation categories. Finally, \sectionname~\ref{sec:Discussion} summarises our main contributions, limitations of this study and final remarks.

\section{Data and networks}
\label{sec:data}

We analyse the emergence of segregated communities in the scientific co-authorship network, focusing on the field of Computer Science. The choice of Computer Science here is pragmatic (manageable size) but also because we can study
co-authorships in this field since its early stages; it
consolidated as a discipline relatively recently (the late 60s) with
the appearance of associations, undergraduate and PhD programmes, and specialised
funding agencies~\cite{mattitedre_2017_the}. We obtained data from
the Semantic Scholar Open Research Corpus~\cite{Lo2020}. Our analyses
correspond to 45 years from 1975 to 2020 for which we have
sufficient data. To
simplify the manuscript, we display some of the main results of our
analysis using one particular year (2010) as an example. The choice of example year is somewhat
arbitrary and was driven solely by the idea that approximately 10
years of work after that year should provide enough information about
citation trends. For generality, we study other 2 example years (2006 and 2014) with results given in the
Supplementary Material. Henceforth, all references to results in the Supplementary Material have
a prefix ``S'' (e.g. \sectionname~S1, \figurename~S1, \tablename~S1). All 3 example years have similar results regarding the structure of
the communities but differ in some of the citation analyses. We leave
a complete longitudinal analysis across all years for future work,
noting that citation comparisons cannot be fairly performed for recent years as works have yet to accrue citations.

For each year of analysis, we build a co-authorship network. Each node represents a researcher. A link is created when 2 researchers co-author at least one scientific publication in the year of study. For the analyses in this paper, we select the Largest Connected Component (LCC) of each co-authorship network. The characteristics of the LCC co-authorship networks for the 3 years studied are shown in \tablename~\ref{tab:Network_metrics}. Values in parentheses represent the proportion of the metric in the LCC compared with the entire co-authorship network. For example, for building the co-authorship network in 2010, we used all of the 615,737 available publications but then just analysed 294,181 publications in the LCCs (0.48 of the available publications). 

\begin{table}[ht]
\caption{\textbf{Characteristics of the Largest Connected Component (LCC) co-authorship network in 2006, 2010, and 2014.} The values in parentheses correspond to the proportion of each quantity falling within the LCC as a fraction of the entire co-authorship network (e.g. for 2010, there were 294,181 papers forming the LCC, which is 0.48 of all Computer Science papers available in that year). The communities were detected with the \textit{Label-propagation} algorithm~\cite{raghavan2007near}. Information about the growth of these metrics per year is given in \sectionname~S1.}

\resizebox{0.98\textwidth}{!}{%
      \begin{tabular}{lrrr}
        \toprule
        Metric per year  & 2006 & 2010 & 2014\\
        \midrule
        Number of papers & 194,114 (0.43) & 294,181 (0.48) & 369,304 (0.52)\\
        Number of nodes & 249,797 (0.47) & 407,532 (0.54) & 566,835 (0.57) \\
        Number of edges &  292,336 (0.22) &  1,453,217 (0.29) &  1,042,623 (0.32) \\
        Density & 9.37e-06 & 1.75e-05 & 6.49e-06 \\
        Clustering coefficient & 0.78 & 0.99 & 0.89 \\
        Mean degree & 4.97 & 13.12 & 6.48 \\
        Mean weighted degree & 5.99 & 14.33 & 9.44 \\
        Mean strength degree & 1.73 & 1.78 & 1.8\\
        Number of communities ($\geq$3 researchers) & 24,470 & 39,998 & 54,655 \\
        Number of researchers in communities ($\geq$3 researchers) & 249,797 & 407,532 & 566,835 \\
        Number of internal papers (all the authors within & 86,354 & 128,415 & 189,072 \\
        the same community) & & & \\
        \bottomrule
      \end{tabular}
      }
      \label{tab:Network_metrics}
\end{table}

There are different ways to measure the value of the links between 2 researchers. For the current analyses, we use the strength of the link between 2 researchers $i$ and $j$ as proposed by Newman~\cite{Newman2004,Cann2019}. The strength captures the idea that 2 researchers that are the sole co-authors of a paper know each other better than 2 researchers that co-authored a paper with many other co-authors, hence giving more importance to those papers with fewer co-authors. The strength is calculated as $w_{ij}=\sum_k\frac{\delta_i^k\delta_j^k}{n_k-1}$, where $\delta_i^k$ takes the value of 1 if the researcher $i$ co-authored the paper $k$ and $n_k$ refers to the number of authors of the paper $k$. To sum the strength of the links of $i$ leads to the strength degree, which differs from the 2 well-known options of giving a value of 1 to each link (leading to the degree) or using the number of co-authorships as the weight of the link~\cite{barrat2004} (leading to the weighted degree). In \tablename~\ref{tab:Network_metrics}, we compare the mean value of the 3 degrees (degree, weighted degree and strength degree) computed for the LCC. In \sectionname~S2, we give toy examples showing how the 3 degrees are calculated and compare their distributions over the years.

\section{Community detection and description}
\label{sec:Community_partition}

To compute the community partition of the entire co-authorship network, we tested 6 commonly used community detection algorithms divided into 2 categories: modularity optimisation (Leading-eigenvector~\cite{Newman-Girvan2004}, Multilevel~\cite{blondel2008fast}, Fast-greedy~\cite{Clauset2004}) and dynamical processes (Infomap~\cite{Rosvall2008}, Walktrap~\cite{Pons2005}, Label-propagation~\cite{raghavan2007near})~\cite{Fortunato2016}. To select which algorithm represents a better community detection, we must
consider that all the co-authors of one publication form a
clique~\cite{Newman2001}, resulting in high clustering coefficients for co-authorship networks~(\tablename~\ref{tab:Network_metrics}). Following the methodology proposed by Fortunato and Hric~\cite{Fortunato2016}, we select the results from the \textit{Label-propagation} algorithm~\cite{raghavan2007near} because it finds communities that are less confounded by fully connected cliques and have higher average embeddedness of their nodes. The embeddedness of a node is its internal (inside the community) strength degree over its total strength degree~\cite{Lancichinetti2010}. The results of each algorithm are included in \sectionname~S3. In addition, we analyse if our results depend on the community partition in \sectionname~S9, where we repeat some of the analyses for communities computed with \textit{Infomap}~\cite{Rosvall2008}.

\begin{figure}[ht]
    \centering
    \includegraphics[width=0.95\textwidth]{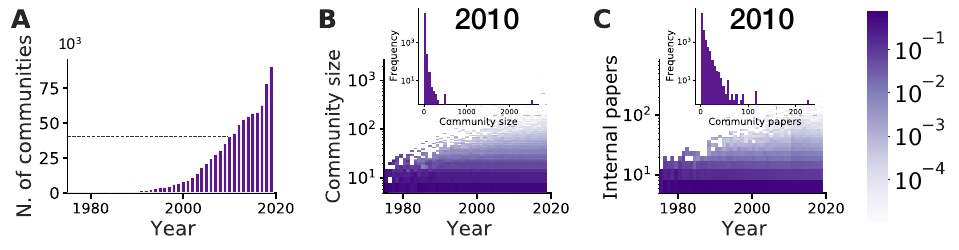}
    \caption{\textbf{Computer Science co-authorship community structure from 1975-2020.} Plots show community metrics based on the \textit{Label-propagation} algorithm for the Largest Connected Component (see text). (A) Number (in thousands) of communities per year. The dashed line highlights the 39,998 communities with size $\geq$3 in 2010. (B) Community size distribution (i.e., number of researchers per community). (C) Distribution of the number of internal papers (all authors within the same community) per community. The colour bar represents the proportion of communities with a given size and a number of internal papers. The inset panels of (B) and (C) show the frequencies of size and number of internal papers for the example year (2010).}
    \label{fig:Communities_size}
\end{figure}

\figurename~\ref{fig:Communities_size} shows the evolution of community structure from 1975 to 2020 based on outputs from the \textit{Label-propagation} algorithm. The number of communities has grown above 50,000 by the end of the study period (\figurename~\ref{fig:Communities_size}A). The distributions of community sizes (number of nodes in each community) for each year are shown in \figurename~\ref{fig:Communities_size}B; the community size frequencies for 2010 are presented in the inset. Interestingly, $90\%$ of the studied communities have fewer than 20 researchers, a constant tendency each year. The maximum community size in the last five years of data (2015-2020) is more than 2,000. Finally, analysing the number of internal papers (i.e. papers with all the authors within the same community) written by each community, we found that $94\%$ of communities publish less than 10 papers, with an upper limit slightly above 200 papers (\figurename~\ref{fig:Communities_size}C). On average, for all the years, there are 0.38 internal papers per researcher (average number of internal papers over the number of researchers). The last results indicate that most researchers in Computer Science work in medium size groups, with the majority working on a few papers, differing from other disciplines with solo authors or large working groups~\cite{Fanelli2016}.

\section{Community segregation}
\label{sec:Communities_segregation}
From this section, all analyses are done considering the researchers, internal papers and communities in the LCC co-authorship network. In addition,  because we study the internal connectivity structure of the communities, we analyse communities with at least 3 researchers. Hence, for 2010, we analysed 128,415 papers authored by 407,532 researchers grouped in 39,998 communities, as shown in the last 3 rows of \tablename~\ref{tab:Network_metrics}. 

\subsection{Spectral segregation index}
\label{sec:Spectral_segregation_index}
We use the Spectral Segregation Index (SSI) proposed by Echenique and Fryer~\cite{Echenique2007a} to measure structural segregation in the detected communities of the LCC. The SSI measures individual segregation as the linear combination of a node's and its neighbours' fraction of internal connectivity inside the group defined (internal refers to links inside the community in our case). The SSI implies a reinforcing process in which a node with a high SSI value has neighbours with a high SSI. There are various segregation metrics, and an interested reader should refer to Bojanowski and Corten~\cite{Bojanowski2014}.


We compute the SSI following the procedure defined by Echenique and Fryer~\cite{Echenique2007a}: First, we normalise the LCC's adjacency matrix $R=[r_{ij}]_{N \times N}$ (which contains the strength of the link between 2 researchers $i$ and $j$). To achieve this, we take the original adjacency matrix and normalise their rows, to sum up to 1 (one). Then, we select a submatrix $B_g$ for each community, $g$, which contains only internal interactions within the community $g$. The value of $SSI_g$ corresponds to the largest eigenvalue $\lambda$ of the submatrix $B_g$~\cite{Echenique2007a}.

The eigenvalue $\lambda$ is computed as the stationary state of a ``random walk'' process. Hence, the connectivity patterns within the community shape the values of $\lambda$, which is, in turn, the average of the individual segregation values within the community. Values of SSI near 0 represent a low segregation level, while values near 1 represent high segregation. Communities that are disconnected components have an SSI equal to 1, meaning perfect segregation~\cite{Echenique2007a}, hereafter referred to as completely segregated communities.

\subsection{Defining segregation categories}
We compute the SSI considering only the connections within a calendar year in the co-authorship network, and we use communities with $\geq$3 nodes, considering only the LCC. For 2010, we worked with 39,998 communities (29\% from the original 136,967 communities of the entire co-authorship network in 2010). The other communities are completely segregated and do not connect to the LCC. Completely segregated communities: {\em (i)} can be cliques (i.e., fully connected subgraphs), {\em (ii)} have few internal papers (1.44 on average),  and {\em (iii)} do not have a core position (computed from the k-core decomposition of the communities network). Their presence is partially due to the time-window considered (i.e., one year); the longer the period considered, the larger the LCC becomes and, consequently, the fewer the isolated components. Then, we do not analyse the structural properties or core positions of completely segregated communities in \sectionname~\ref{sec:Communities_topology} because they could skew our results. However, we include a category of completely segregated communities in \sectionname~\ref{sec:Citations_researchers} when we analyse the relationship between different segregation levels and citations.

\begin{figure}[ht!]
    \centering
    \includegraphics[width=0.85\textwidth]{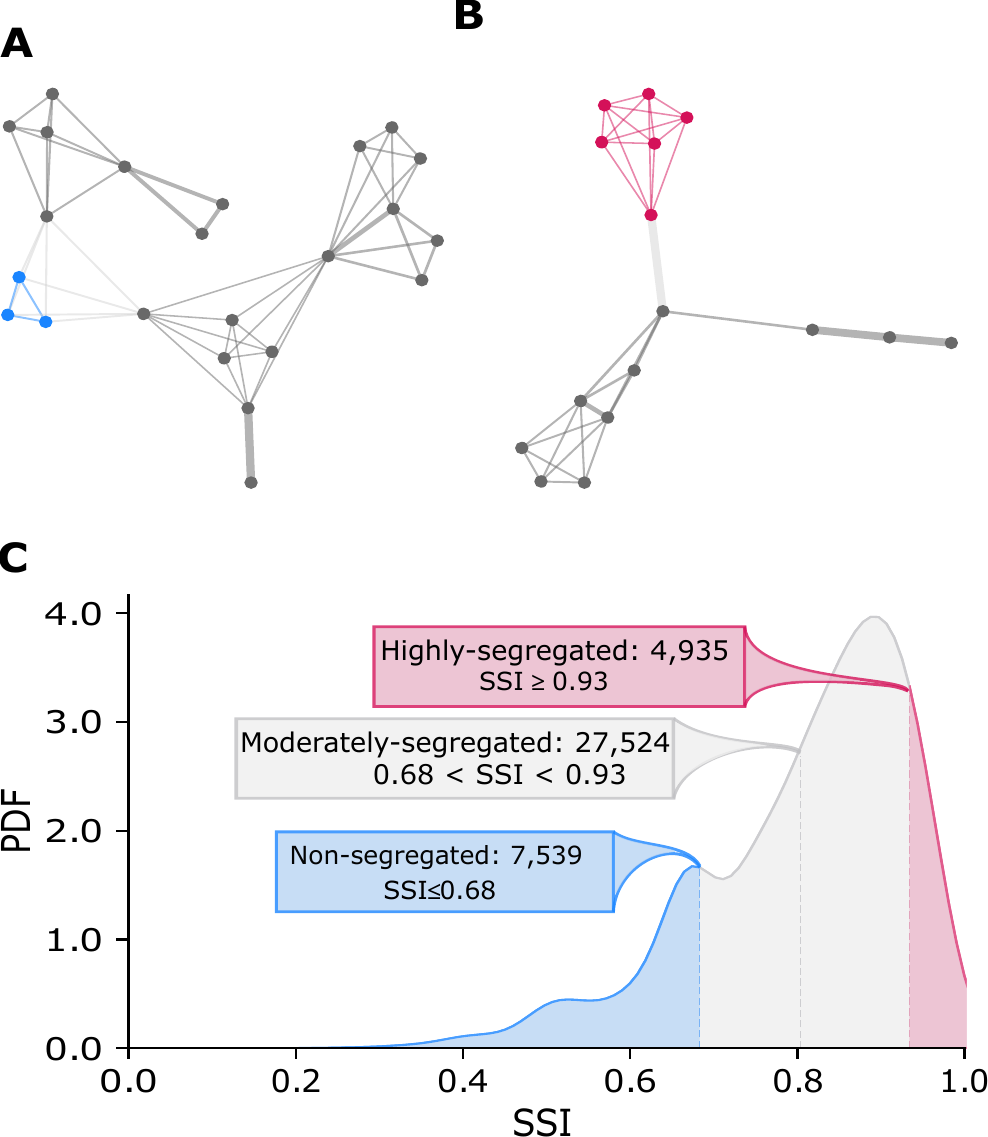}
    \caption{\textbf{Classifying communities as non-segregated and highly segregated.} (A) and (B) are examples of ego-networks of co-authorships in 2010 of non-segregated and highly segregated communities, in light blue and red, respectively. Ego-networks are sub-graphs induced by the connections between central nodes, i.e., ego (colored nodes belonging to the selected community) and their one-step neighbours, i.e., alters (dark grey nodes belonging to other communities connected to the colored community). Edges inside the communities have the color of the nodes, while links across communities are in light grey. (C) shows the probability density function (PDF) of the spectral segregation index (SSI) for 2010. The plot is divided into 3 categories that denote non-segregated (light blue), moderately  (grey), and highly segregated (light red) communities. The complete procedure is in \sectionname~\ref{sec:Spectral_segregation_index}. For the distribution, we use a Gaussian kernel density estimation with the “rule of thumb” for the bandwidth selection~\cite{scott2015multivariate}.}
    \label{fig:Partitions_SSI}
\end{figure}

The values of SSI are continuous, and there are no clearly defined categories, so we developed a procedure to identify ordered categories. First, we compute the probability density function PDF of the SSI, its mean ($\mu$) and standard deviation ($\sigma$). Second, we select as highly segregated those communities with a relatively high $\textnormal{SSI}\geq\mu+\sigma$, and non-segregated those communities with a relatively low $\textnormal{SSI}\leq\mu-\sigma$. This approach naturally leads to 3 categories of segregation: highly, moderately, and non-segregated. In \figurename~\ref{fig:Partitions_SSI}C, we show the PDF of SSI for 2010, the division of segregation categories, and the number of communities in each category. This procedure ends with 7,539 non-segregated,  27,524 moderately segregated, and 4,935 highly segregated communities. We compute the same analysis in \sectionname~S4 for 2006 and 2010.

In~\figurename~\ref{fig:Partitions_SSI}, we show toy networks of non-segregated and highly segregated communities in panels~A~and~B, respectively. Those toy networks show communities with their members in colour, grey for nodes from other neighbouring communities and in light grey links among different communities.

In the following analyses, we concentrate on studying 2 categories: non-segregated and highly segregated communities, as we want to study the extremes of the SSI spectrum. However, in the first subsection of \sectionname~\ref{sec:Citations_researchers}, we compare the citation patterns of the 4 ordered segregation categories: completely segregated, highly segregated, moderately segregated and non-segregated communities.

\section{Characterisation of communities in different segregation categories}
\label{sec:Communities_topology}
We compare 4 metrics in total to investigate the characteristics of non-segregated and highly segregated communities. The first 3 metrics refer to the structural properties of the communities to understand if the segregation categories are related to a community's internal connections. We compute the size (measured as the number of researchers), density (measured as the proportion of internal links over the set of all possible internal links), and clustering coefficient (measured as the number of triangles over the number of triplets within the community)~\cite{Newman-Girvan2004}.

The fourth metric refers to the core position of the communities because the core/periphery position of segregated communities in online social networks (i.e. echo chambers)~\cite{Williams2015} has been shown to influence their ability to spread information during social movements~\cite{Barbera2015}. Therefore, in the context of scientific production, we want to understand if the communities' position in the co-authorship network also relates to their segregation category. We first create a network in which each community is a node, and links between these nodes exist if their members share co-authorships. Then, we apply the k-core decomposition algorithm~\cite{Batagelj2003} and assign each community to a correspondent core. The core values range from 1 (periphery) to N (nucleus), where N depends on how many cores we have in a particular year, 11 in the case of 2010. See \sectionname~S5 for more details about calculating the core decomposition of the communities networks.

\begin{figure}[t]
    \centering
    \includegraphics[width=0.95\textwidth]{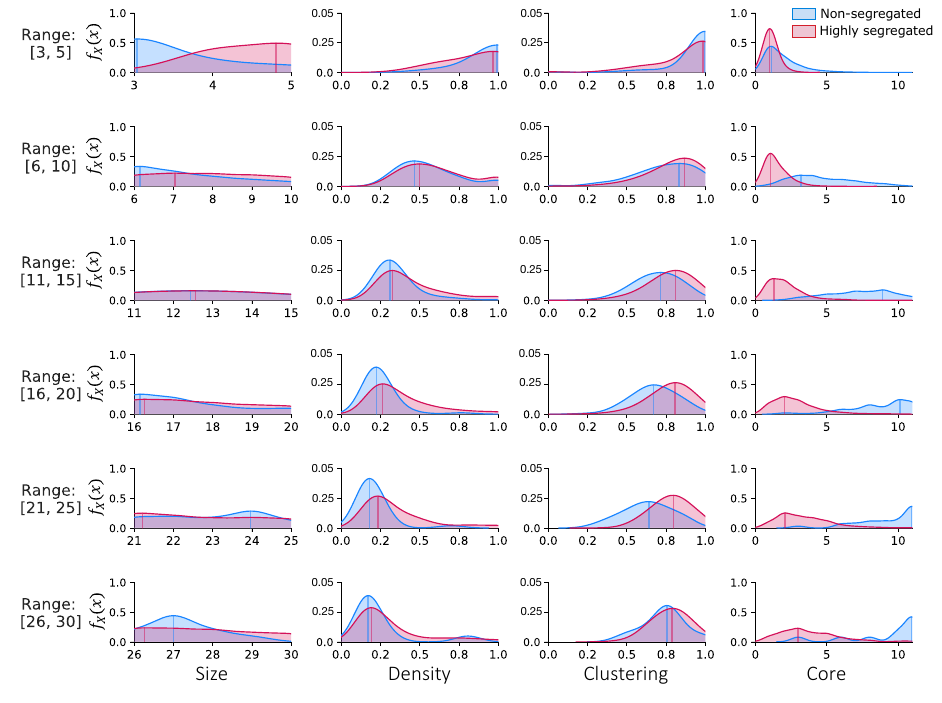}
    \caption{\textbf{Topological and core position differences among non-segregated and highly segregated communities.} The panels represent the probability density functions (PDF) in each column for the size, density, clustering, and core position of highly segregated (light red) and non-segregated (light blue) communities for different size ranges per row. When plotting the curves, we use a Gaussian kernel density estimation with the ``rule of thumb'' for the bandwidth selection~\cite{scott2015multivariate}.}
    \label{fig:Topological_metrics}
\end{figure}

As a previous step, we group the communities by different size ranges (detailed explanation and more analyses in \sectionname~S6). For the comparison, we first separate the communities by size range and segregation category (i.e., highly or non-segregated). Then, we perform a statistical analysis to compare the PDF of the 4 metrics (size, density, clustering, and core position) of the non-segregated and highly segregated communities, with results for the 2010 network in \figurename~\ref{fig:Topological_metrics} and analogous plots for different years in \sectionname~S7. The sixth range of communities' size shown in \figurename~\ref{fig:Topological_metrics} goes up to 30 as this value is the largest communities' size where there are at least 30 non-segregated communities. The last suggests that it is difficult for large communities to be non-segregated.

Our results show that for small communities, there are no differences between non-segregated and highly segregated communities in terms of density, clustering or core position. However, as the communities grow, the density column shows both types of communities decreasing their peak values from 1 to 0.2. The clustering column shows decrements from 1 to 0.5 for non-segregated and 0.8 for highly segregated communities. We infer that these decrements are expected for larger communities, as they can be formed by different groups with enough intergroup co-authorships. For the core position, there are no differences when communities are smaller than 5, with both types being in the periphery. However, when the size increases, on one side, non-segregated communities start to be in higher-value cores until the largest ones reach the nucleus. And on the other side, highly segregated communities remain in peripheral cores.

To highlight the importance of disaggregating communities by size, we perform the same analyses without separating them by size in \sectionname~S7.1. The results are indeed misleading when communities with different size ranges are mixed, as the number of nodes and links affect density and clustering and hide the differences in the core position.

In conclusion, small communities tend to be denser, more clustered and toward the network's periphery. As expected, their densities and clustering decrease when they increase in size, though less visibly for clustering. There are mild differences in density and clustering between non-segregated and highly segregated communities, with values mainly driven by community size. Moreover, there is a difference in their core position, with more large non-segregated communities in the nucleus and more highly segregated communities in peripheral cores.

We performed 3 additional analyses, reported in the Supporting Materials: \textit{i}) We repeat this analysis for the years 2006 and 2014 in \sectionname~S7.1 with similar results: The communities located in the network's periphery are more numerous and smaller, and those highly segregated in the nucleus have larger sizes. \textit{ii}) We compare the Z-Score of the metrics, and compute kernel density estimators for comparing size, SSI, and core position at the same time in \sectionname~S7.2. The results remain congruent with statistical differences between non-segregated and highly segregated communities for the core position when communities are large. \textit{iii}) We repeat the procedures of this section in \sectionname~S9.2 with the results of \textit{Infomap}. We found that both algorithms have similar results. However, \textit{Label-propagation} always has more highly segregated researchers than non-segregated (if we use the same characterisation we have done here), while it changes for \textit{Infomap}: there are more highly segregated researchers in the periphery and more non-segregated researchers in the nucleus.

\section{The effect of segregation on citations}
\label{sec:Citations_researchers}
This work's third and final research question relates to understanding the relationship between segregation and citation levels. Citations are a well-known measure of scientific success, but we also encourage reading our results critically, as citations have been related to selection biases, mainly affecting underrepresented communities, e.g. women and non-western researchers publishing non-English content~\cite{Cronin2015}. Here, we consider both the number of citations and the origin of the citations (in terms of the community partition) to characterise whether highly segregated communities have more self-citations than non-segregated ones. For each researcher in non-segregated and highly segregated communities, we analyse the citations received until 2020 by the publications of 2010.

First, we investigate whether the number of internal papers correlates with \textit{i}) the total number of citations and \textit{iii}) the average number of citations per paper as in previous literature, the number of citations an author receives has been related to their number of publications~\cite{Huang2020}. We find low correlations of 0.29 ($p$-value $<$$10^{-3}$) and 0.10 ($p$-value $<$$10^{-3}$), respectively. We use the Spearman correlation in both cases because the number of papers has a non-linear relationship with citations and citations per paper (\figurename~S13).

Second, we compute the cumulative density function CDF of 4 variables for researchers within the specific category of communities: {\em (i)} total number of citations, {\em (ii)} citations per paper, {\em (iii)} proportion of citations from the same community, and {\em (iv)} proportion of all citations from the same year's co-authors (2010 for the main manuscript).

\begin{figure}[ht]
    \centering
    \includegraphics[width=0.95\textwidth]{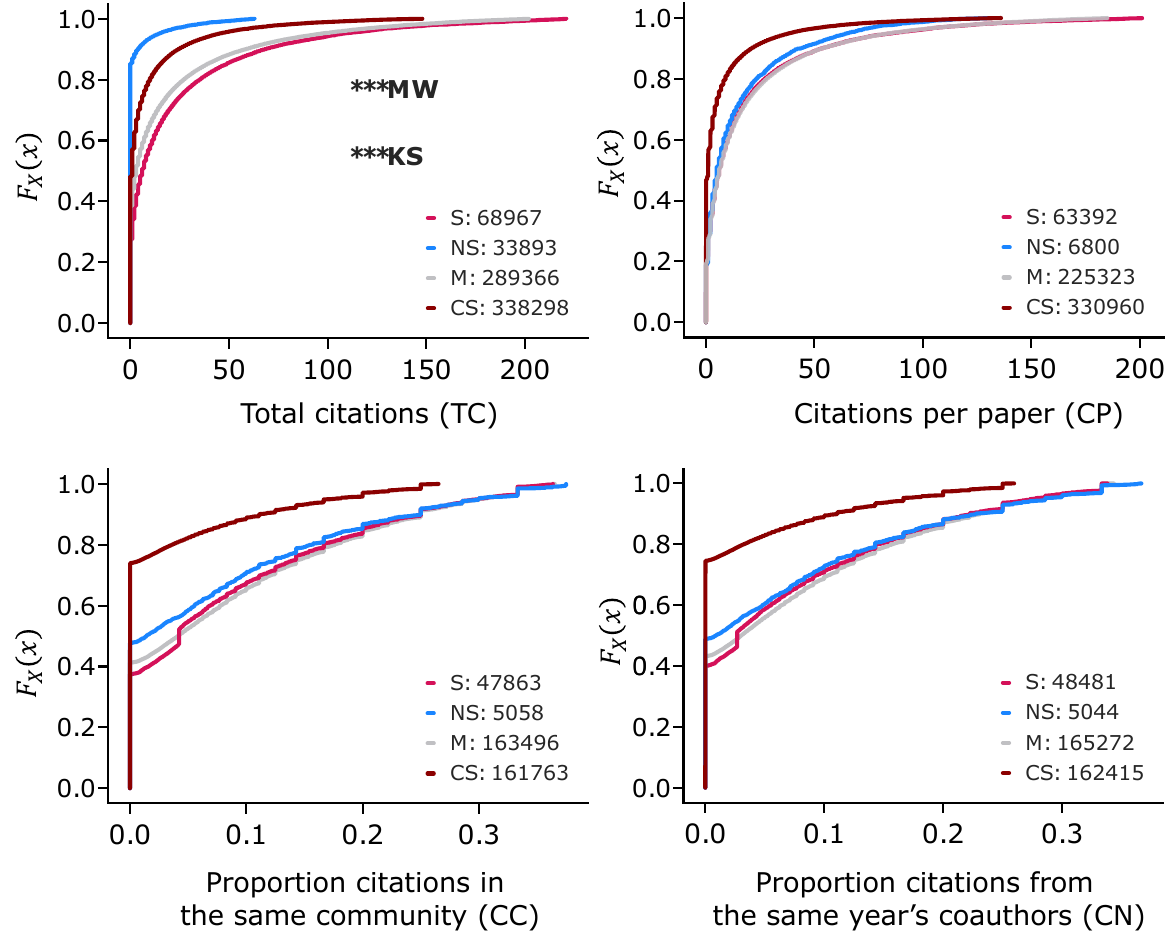}
    \caption{\textbf{Citation metrics for all researchers in communities of different segregation categories.} Each panel represents the cumulative density function (CDF) for the total citations (TC), the citations per paper (CP), the proportion of citations from the same community (CC), and the proportion of citations from the same year's co-authors (CN). The code of colours is: dark red for researchers in completely segregated (CS), grey for moderately segregated (M), light red for highly segregated (S), and blue for non-segregated communities (NS). Letters \textbf{KS} or \textbf{MW} appear when there are significant $p$-values for Kolmogorov-Smirnov (different distribution shapes) and Mann-Whitney (different distribution medians) for the CDFs of non-segregated and highly segregated communities. Significance levels are denoted as follows: * $<$ 0.1, ** $<$ 0.05, and *** $<$ 0.01.}
    \label{fig:Citations_plot_all}
\end{figure}

\begin{figure}[ht]
    \centering
    \includegraphics[width=0.95\textwidth]{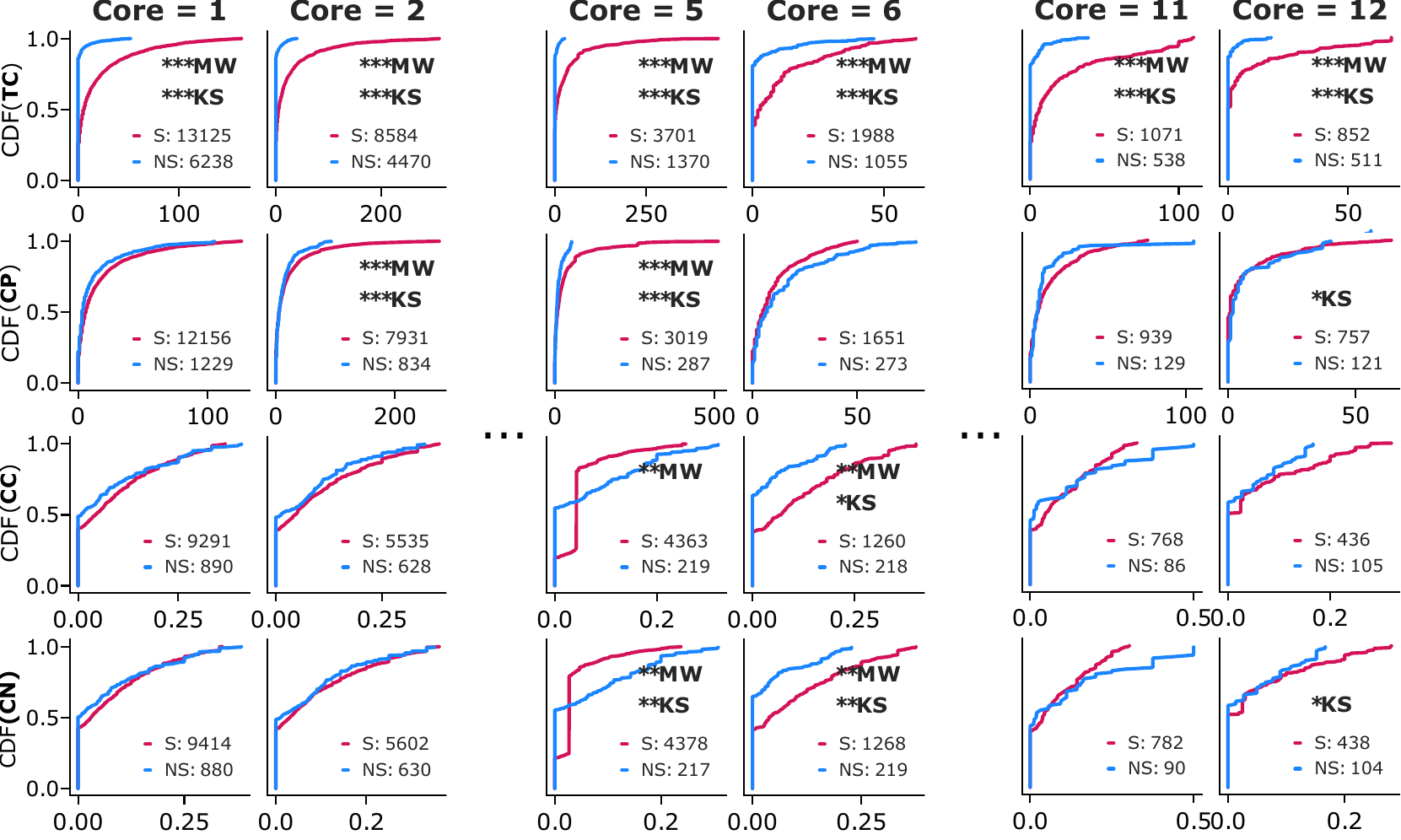}
    \caption{\textbf{Citation metrics for researchers in communities of different segregation categories and core positions.} Each row represents the cumulative density function (CDF) for the total citations (TC), the citations per paper (CP), the proportion of citations from the same community (CC), and the proportion of citations from the same year's co-authors (CN). The code of colours is: light red for highly segregated (S) and blue for non-segregated communities (NS). Letters \textbf{KS} or \textbf{MW} appear when there are significant $p$-values for Kolmogorov-Smirnov (different distribution shapes) and Mann-Whitney (different distribution medians) for the CDFs of non-segregated and highly segregated communities. Significance levels are denoted as follows: * $<$ 0.1, ** $<$ 0.05, and *** $<$ 0.01. Here, we show 7 out of 11 cores to guide the reader, but \figurename~S15 shows results for the 11 cores of 2010.}
    \label{fig:Citations_plot_cores}
\end{figure}

For each variable, we analyse researchers at 2 levels of granularity. {\em (i)} All researchers without grouping them by core position for the 4 categories:  completely segregated, highly segregated, moderately segregated and non-segregated in \figurename~\ref{fig:Citations_plot_all} (definition of each category in \sectionname~\ref{sec:Communities_segregation}), and {\em (ii)} researchers grouped by the core position of their communities for 2 categories: non-segregated and highly segregated in \figurename~\ref{fig:Citations_plot_cores}. We did not analyse our results by different ranges of internal papers due to the low correlation with the citation variables.

We use 2 statistical tests to compare the CDFs of non-segregated and highly segregated communities: Kolmogorov-Smirnov (KS) and Mann-Whitney (MW). The first test compares the shape of the distributions, and the second compares the differences between medians.


We first analyse the CDFs for the {\em (i)} total number of citations TC and {\em (ii)} citations per paper CP. On an aggregated level, in \figurename~\ref{fig:Citations_plot_all} top row, our results indicate that highly segregated researchers have more TC than non-segregated researchers. Considering the number of CP, we see that completely segregated researchers (darker red in the plot) have smaller values than other researchers, with no significant differences. However, the previous results hide some information because they are averaging over all network cores. Then, in \figurename~\ref{fig:Citations_plot_cores}, we group the researchers by the core position of their communities, and we split the results into the nucleus, middle, and periphery. In middle and periphery cores, highly segregated researchers have more TC than non-segregated ones, with opposite results in the nucleus (top row). For the CP (second row), there are no differences in the middle or periphery cores, but non-segregated researchers have more CP in the nucleus.

The results of TC and CP in 2010 are similar in 2006 and 2014. For TC, highly segregated researchers outperform non-segregated in the periphery and middle cores, but there are no significant differences for CP. In the nucleus, for both TC and CP, non-segregated researchers do better (detailed results of 2006 and  2014 in \sectionname~S8).

Then, we analyse the CDFs for {\em (iii)} the proportion of citations from the same community CC and {\em (iv)} the proportion of citations from the same year’s co-authors CN. For computing these proportions, we count the number of publications with at least one of the authors in the citing publication satisfying the rule of being in the same community (for CC) or co-author (for CN, regardless of the community). Then, we divide these counts by the total number of citations.

On an aggregated level (\figurename~\ref{fig:Citations_plot_all} second row), our results show no statistically significant differences when researchers are in highly or non-segregated communities. However, completely segregated researchers (darker red) receive lower CC and CN than others. When we group by the core position (\figurename~\ref{fig:Citations_plot_cores} third and fourth rows), there are no differences in the periphery. However, in middle cores, highly segregated researchers have more CC and CN, and in the nucleus, non-segregated researchers have larger values. When we compare these results with the other years, for 2006, there are no differences in CC and CN for non-segregated and highly segregated researchers, but for 2014 the trends are similar to those in 2010 (\sectionname~S8).

In summary, highly segregated researchers tend to have more citations per paper when they locate in peripheral cores and more citations from their communities in middle cores. At the same time, non-segregated researchers show higher values for the 4 metrics when they are in cores near the nucleus.

\section{Discussion}
\label{sec:Discussion}
Due to a range of social mechanisms, processes, and biases, co-authorship networks are organised in communities~\cite{Newman2006}. Within-group dynamics might lead to the emergence of segregation and polarisation, hampering innovation, social learning, and problem-solving~\cite{Kim2019,Sasahara2021,Perra2019,Henry2011}. Nevertheless, cohesive groups allow for the development of common narratives and language, offer support and share knowledge. As such, they have been identified as a locus for exploitation (when large in central locations) and exploration (when small in the periphery) of ideas, results, and methods~\cite{Painter2021, Wu2019}. Still, understanding segregated groups in co-authorship networks and their possible effects is limited. Here, we tackle this problem by quantifying segregation levels of communities in co-authorship networks and characterising their topological properties and position in the network.

For our case study, we analyse the co-authorship network of Computer Science in the Semantic Scholar Open Research Corpus~\cite{Lo2020}. We detect communities with the \textit{Label-propagation} algorithm and compute a structural segregation metric considering the community's links: the Spectral Segregation Index (SSI). Based on the distribution of the SSI, we identify 3 main categories and focus on just the 2 opposite limits: non-segregated and highly segregated communities. Then, we compare the communities' size, density, clustering, and core position between categories. Furthermore, we study the relationship between segregation and impact using citations from the community's publications.

Our results indicate that highly segregated communities tend to be more on the periphery, with some differences in density and clustering with non-segregated communities. This finding aligns with previous results~\cite{Malvestio2020}, where the k-core structure of some empirical and randomised networks were shown to be explained by their community structure. When we analyse the total number of citations, researchers in highly segregated communities receive more citations than non-segregated ones in middle and peripheral cores. In addition, when we analyse the sources of those citations, for researchers in highly segregated communities, up to 5\% more of those citations come from the same community than non-segregated communities in middle cores. Combining both results and based on previous literature, we speculate that in terms of spreading ideas and knowledge in the co-authorship network: {\em (i)} researchers in highly segregated communities attract more citations in the periphery of the network because most cited papers are not the internal ones but rather those across communities with diverse disciplines and co-authors~\cite{Zingg2020}. And {\em (ii)} researchers in non-segregated communities in the nucleus are citing themselves more and are exploiting/echoing scientific research ~\cite{Mason2012}.

Both effects need further analysis because, as expected, highly segregated communities located on the periphery have a larger impact. Individual success correlates with the exploitation of ideas~\cite{Mason2012}, but also the most innovative research (exploration of new concepts and persistent citations) comes from the periphery of networks~\cite{Painter2021}, and it is done by smaller groups of researchers~\cite{Wu2019}. Here, our results align with previous evidence showing nodes in the periphery being less active~\cite{Williams2015} (i.e. publishing less in our case) but having more impact. In addition, researchers in those communities are a large population that could become a collective power that can mobilise and spread information~\cite{Barbera2015} (such as scientific theories).

Researchers in larger and non-segregated communities in the nucleus also increase their impact. These results need further exploration because their central positions in the network's nucleus increase their chance of outside interactions with highly segregated communities, which can accelerate the propagation of echoed information (ranging from biased theories to new paradigms) from local groups to reach the entire network~\cite{Davis2020}. The inner impact of highly segregated communities and their impact on the whole network should be measured to intervene, if necessary, and tackle or boost the spread of echoed information to different groups~\cite{Jalali2020}.

\subsection{Limitations}

First, our analysis does not generalise for all the years of Computer Science papers available in the Semantic Scholar database because we study just 3 years. We have developed a repeatable methodology and replicated our findings over several years. Still, further analysis is needed to understand how the transitions of researchers between different segregation levels affect their research impact over time.

Second, our analyses only generalise to some co-authorship networks because the publications of Computer Science in the Semantic Scholar Open Research Corpus represent a vast amount of literature in a discipline prone to working in small teams~\cite{Newman2001}. Further analysis of other fields is needed to understand how these patterns apply to different co-authorship structures.

Third, we did not classify the core-periphery type of our network. Recent work has highlighted the importance of understanding if the network is prone to be divided into cores as layers (as we did with the k-core decomposition algorithm) or if a hub/spoke core division is a better descriptor~\cite{Gallagher2021}. However, their results show that authorship networks are the most prone to have a core-layered typology, as we used in the current work. In further analyses, the definition of segregated communities should also consider the co-authorship network's core typology.

Finally, our fourth limitation relies on using the extreme values of the SSI's PDF from the co-authorship networks to define segregation categories of communities. A more precise analysis could consider continuous values of the SSI, other features and data to represent better the consumption and production of scientific knowledge~\cite{Zeng2017}. Future work could consider a continuous comparison of the metrics used in this analysis, publications' content, researchers' demographic diversity, and interdisciplinary citations.

\subsection{Future research}

Future research on this topic could consider: {\em (i)} the temporal analysis of segregated communities and their relation to gaining more or fewer citations over time, {\em (ii)} the analysis of the diversity of the scientific publications inside the communities using opinion distance~\cite{Sasahara2021} and their demographic diversity to understand if the segregated and isolated communities are not diverse and echoing research to the point of becoming polarised, {\em (iii)} the definition of lead researchers (using the hub/spoke core or author position in the publications) and the understanding of their relationship to segregated communities~\cite{Guo2020}, iv) the measurement of the impact of segregated communities on the topology of the network formation and the spreading processes of scientific theories~\cite{Tornberg2018}.


\section*{Declarations}

\subsection*{Availability of data and materials}
The datasets generated and/or analysed during the current study are available in the Semantic Scholar repository, \url{https://www.semanticscholar.org/product/api}

\subsection*{Competing interests}
The authors declare that they have no competing interests.

\subsection*{Author's contributions}
All authors conceived and designed the research. AMJ acquired the data. AMJ, HTPW, NP and RM analysed the data. All authors discussed the research, wrote and approved the final version of the manuscript.

\subsection*{Acknowledgements}
The authors would like to thank the U.S. Army Research Office for the partial
support provided to RM under grant number W911NF-18-1-0421. AMJ is funded by a PhD studentship from the UK Engineering and Physical Sciences Research Council. No funding bodies had any influence over the content of this report.

\bibliographystyle{abbrv}








  \subsection*{Additional file 1 --- Supplementary material, including details on methods used in this research.}


\end{document}



\setlength{\abovedisplayskip}{10pt}
\setlength{\belowdisplayskip}{10pt}

\maketitle

This text contains additional material with in-depth analyses to better explain the results of the Main Manuscript. The Main Manuscript describes the procedures and analysis of the results using the year 2010. This text includes methods and a comparison of the main results with the 2 extra years: 2006 and 2014 to understand whether the results are particular to the chosen year. The selection of these 3 years was somewhat arbitrary; 2010 was selected because it allowed us 10 years of citations. We then decided to use 2 other periods, and we went with a choice of $\pm$4 years. Further analysis of the temporal aspects of being part of highly segregated communities is left as future work. 

Furthermore, the results in the main text refer to communities found using the \textit{Label-propagation} algorithm. To understand whether the results are dependent on \textit{Label-propagation}, we conduct some of our analysis on community segregation, topology, and citations with communities found with \textit{Infomap}. 

This suplementary material is organised as follows. Section~\ref{sec:metricsgrowth} analyses the temporal behaviour of 7 metrics related to the structural properties of the co-authorship networks. Section~\ref{sec:degree_distribution} defines the 3 possible ways to give value to the links of the co-authorship network with a toy example, and it compares the resulting degrees distributions over time. Section~\ref{sec:community_detection_algorithms} describes the 6 community detection algorithms used in this study and the analyses done to decide on a chosen algorithm for the primary analyses. Section~\ref{sec:segregated_communities_other_years} has the results for dividing communities into the segregation categories for the 2 years of comparison, 2006 and 2014. Section~\ref{sec:Network_communities_plot} details the procedure to build the communities' networks and compute their core location. Section~\ref{sec:PDFS_SSI_range_sizes} describes the division of communities by size ranges as a previous step before comparing the structural properties of the communities. Section~\ref{sec:Comparing_Structural_metrics} shows structural and core position metrics analysis over the 3 years with/out dividing by size ranges, and Z-Score analyses controlling by the size of communities. Section~\ref{sec:Citations_communities} analyses the citation patterns of researchers in different segregation categories for the studied years. Finally, Section~\ref{sec:Infomap_analysis} compares the results of our analysis between communities detected with \textit{Label-propagation} versus \textit{Infomap}. Henceforth, all references to the Main Manuscript do not have any letter before the number (e.g., \sectionname~1, Table~1) while the references to this text are prefixed with an S (e.g., Section~\ref{sec:metricsgrowth}).
\section{Co-authorship network metrics}
\label{sec:metricsgrowth}
\begin{figure}[ht!]
    \centering
    \includegraphics[width=0.95\textwidth]{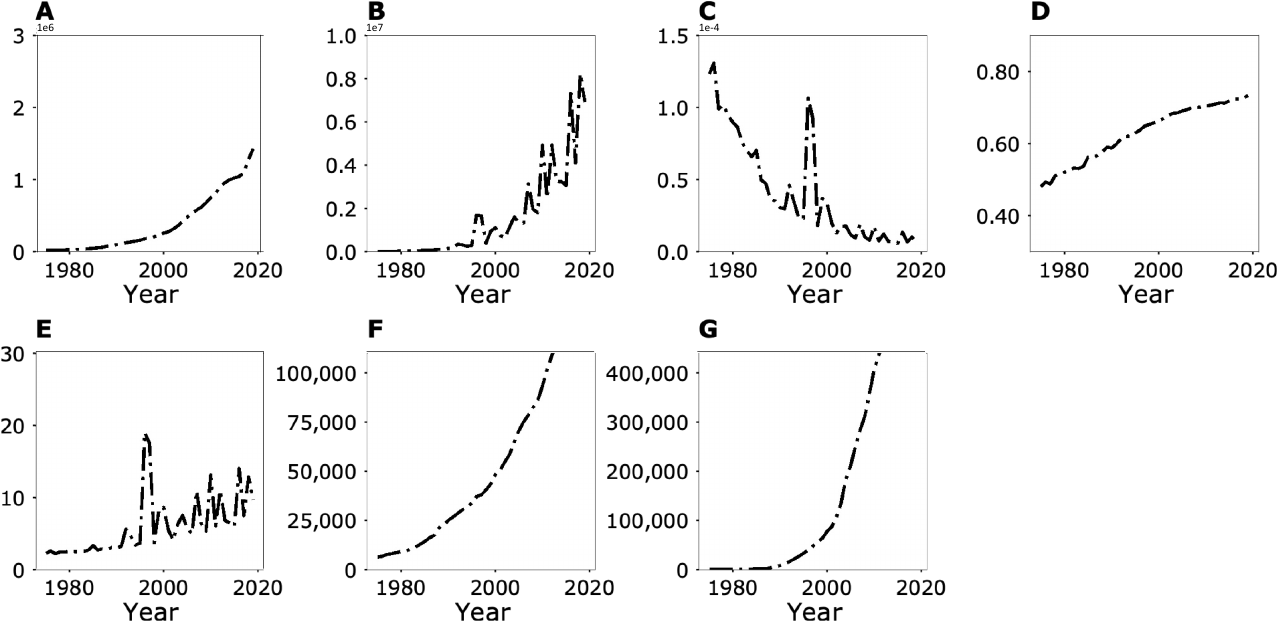}
    \caption{\textbf{Network metrics over the years of analysis.} (A) Number of nodes, (B) Number of edges, (C) Density, (D) Average clustering coefficient, (E) Mean degree, (F) Number of connected components and (G) Largest connected component}
    \label{fig:Temporal_metrics}
\end{figure}

This section analyses the temporal behaviour of variables displayed in \tablename~1.
The order of the panels from A to G has the same order of variables in the table: from ``Number of nodes" to ``Size of the largest connected component". 
We can see in \figurename~\ref{fig:Temporal_metrics}A how the number of nodes grows over time, as well as the number of edges (\figurename~\ref{fig:Temporal_metrics}B). However, as expected, the number of edges growth is not as fast as the number of possible edges $((N \times (N-1))/2)$ because not all new researchers (nodes) can collaborate (get connected) with the already existing ones. Hence, the density decreases over the years with a high peak before the 2000s (\figurename~\ref{fig:Temporal_metrics}C). We were unable to identify the reason for such peak but it is probably due to some database change or addition done at that period.

Co-authorship networks have the particular characteristic of forming fully connected cliques among all the authors of one paper. Then, their clustering coefficient value is expected to be relatively high, considering the network size. In addition, there have been new trends in scientific practices to work in larger teams over the last decades. Then, the clustering coefficient has been increasing over time but with slower increments in the last 10 years (\figurename~\ref{fig:Temporal_metrics}D), increments in the number of papers per community (as seen in \figurename~1C), and a higher connectivity per researcher with the mean degree increase over the years (\figurename~\ref{fig:Temporal_metrics}E). Finally, this increased connectivity of researchers publishing in Computer Science has, on one side, incremented the number of connected components (\figurename~\ref{fig:Temporal_metrics}F), and on the other side, has increased at a faster pace the number of nodes in the Largest Connected Component after the 2000s (\figurename~\ref{fig:Temporal_metrics}F).

\begin{table}[ht!]
\caption{\textbf{Characteristics of the co-authorship network in 2006, 2010, and 2014.} The communities were detected with the \textit{Label-propagation} algorithm. Detailed growth of some of these metrics per year in Section~\ref{sec:metricsgrowth}.}
\resizebox{0.7\textwidth}{!}{%
      \begin{tabular}{lrrr}
        \toprule
        Metric per year  & 2006 & 2010 & 2014\\
        \midrule
        Total number of papers & 446,420 & 615,737 & 710,567 \\
        Total number of nodes & 531,113 & 750,711 & 998,211\\
        Total number of edges & 1,321,184 & 4,926,882 & 3,232,835\\
        Density & 9.37e-06 & 1.75e-05 & 6.49e-06 \\
        Average clustering coefficient & 0.78 & 0.99 & 0.89 \\
        Number of connected components & 75,168 & 93,434 & 118,074\\
        \bottomrule
      \end{tabular}
      }
      \label{tab:Network_metrics}
\end{table}

\newpage 
\section{Analyses of degrees types}
\label{sec:degree_distribution}

\begin{figure}[ht!]
    \centering
    \includegraphics[width=0.95\textwidth]{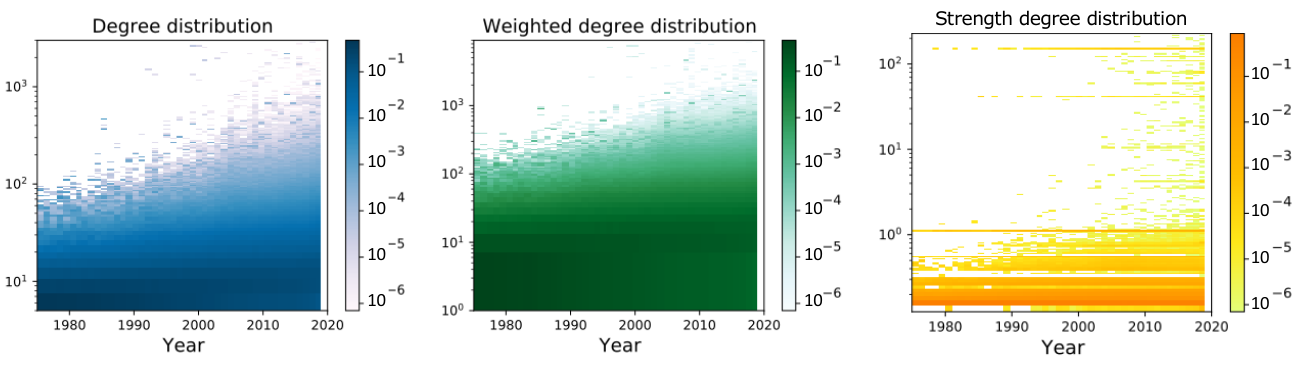}
    \caption{\textbf{Comparison of 3 types of degree distributions based on links weight.} Degree, weighted degree, and strength degree distribution of the co-authorship networks per year. The darker the colour, the higher the frequency of that degree value (y-axis) in the specific year (x-axis). The colour bars of each panel show correspondence between the colour tone and the value of the density of nodes with that degree.}
    \label{fig:Degree_distributions}
\end{figure}

This section shows a toy network in \tablename~\ref{tab:toy_netwroks_degrees} to explain the types of degrees we use and hence understand the distributions over time (\figurename~\ref{fig:Degree_distributions}) for the 3 possible values of the links in co-authorship networks, as mentioned in Section~2.
In the first column of \tablename~\ref{tab:toy_netwroks_degrees}, there is an example of how we are building our co-authorship network: from a list of papers, we select their authors (represented by letters), and we connect those authors with a link if they co-authored a paper. Then, we could have 3 options to give value to those links: degree, which measures the number of co-authors that a researcher is connected to; weighted degree, which measures the number of co-authorships of a researcher which is the sum of the weights of each link (the number of co-authorships between two researchers~\cite{barrat2004}); and strength degree, which measures the relative importance of the co-authorships taking each paper and dividing it by the number of authors minus 1~\cite{Newman2004}.

\begin{table}[ht!]
\caption{\textbf{Toy networks showing the 3 types of degrees.} The first column shows a bipartite representation of 3 papers co-authored by the nodes on the right and how the co-authorship network is built from it. We assigned a symbol to each paper: $\bigtriangleup$, $\square$, and $\bigcirc$ to help the reader. Then, in the co-authorship network, the links are marked according to the corresponding paper. In the last column, the value of each link is followed by the paper that it corresponds to.}
\resizebox{\textwidth}{!}{%
    \begin{tabular}{ccccc}
        \toprule
         Toy network & Node & Degree & Weighted degree & Strength degree  \\
         \midrule
         \multirow{8}{*}{\raisebox{-\totalheight}{\includegraphics[width=0.27\textwidth] {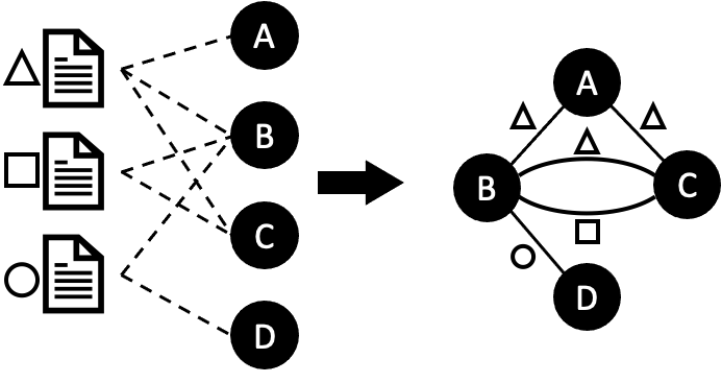}}}& & & & \\
          & A & $2$ & $2$ & $\frac{1}{2}(\bigtriangleup)+\frac{1}{2}(\bigtriangleup)=1$\\
          & & & & \\
          & B & $3$ & $4$ & $\frac{1}{2}(\bigtriangleup) + \frac{1}{2}(\bigtriangleup)+1(\square)+1(\bigcirc)=3$ \\
          & & & & \\
          & C & $2$ & $3$ & $\frac{1}{2}(\bigtriangleup) + \frac{1}{2}(\bigtriangleup)+1(\square)=2$ \\
          & & & & \\
          & D & $1$ & $1$ & $1(\bigcirc)$ \\
          & & & & \\
          \bottomrule
    \end{tabular}
    }
    \label{tab:toy_netwroks_degrees}
\end{table}

In \figurename~\ref{fig:Degree_distributions}, we can see how the distributions of degree (in blue) and weighted degree (in green) grow exponentially over the years, while in the case of the node strength degree (in orange), the growth is more linear with some higher values consistently increasing over the last 10 years. Because we want to understand which co-authorship networks make more segregated communities, we use the strength of the value of the links. If we consider the strength degree, co-authors with high values would have more peer-to-peer interactions for writing a paper, and their values are consistent for more years. This decision is taken to have stronger co-authorships that have fewer co-authors. The example of the toy network in \tablename~\ref{tab:toy_netwroks_degrees} combined with the real values of the distributions in \figurename~\ref{fig:Degree_distributions} show how the number of co-authors has grown over the years, but because the number of papers that each author in Computer Science publishes grows as well, the strength degree remains more or less constant over time.

\section{Comparison of community detection algorithms}
\label{sec:community_detection_algorithms}
We analyse the community structure of co-authorship networks to understand the formation of groups that could be closed in nature,  as mentioned in Section~3.
We applied 6 community detection algorithms divided into 2 categories to avoid biased results in this analysis: optimisation-based and dynamical processes~\cite{Fortunato2016}.

\begin{description}
    \item [Optimisation-based:]
    These algorithms perform optimisation techniques related to the modularity measure, which compares the number of edges within a community with the number of edges expected by chance.
    \begin{description}
        \item [Leading eigenvector:]
        This algorithm expresses modularity in terms of the eigenvectors and performs spectral partitioning for community detection on the modularity matrix~\cite{Newman-Girvan2004}.
        \item [Multilevel:]
        This algorithm maximises modularity in two phases. First, each node is the sole member of a community; then, each node is grouped with each neighbour community, and it computes the modularity again. If the new modularity is larger, the grouped nodes form a new community; otherwise, the new group is discarded. In the case of a local maximum, i.e., no positive gains in modularity, the second phase converts each node in a community with weighted edges with the number of common edges and starts the first phase again. Finally, iterate both phases until achieving a maximum modularity~\cite{blondel2008fast}.
        \item [Fast greedy:]
        This algorithm maximises modularity in faster ways. First, assign each node to a sole member community, and compute a matrix with the gaining in modularity between each pair of communities. Second, find and select the maximum gain in the matrix, merge both communities, and update the gaining in the modularity matrix with the new communities. Finally, repeat the second step until one community remains~\cite{Clauset2004}.
    \end{description}
    \item [Dynamical processes:]
    These algorithms perform flow-based approaches in which the state of the nodes changes as a function of the neighbours' states following spreading process dynamics, with information of the links involved in the community detection as flow paths. 
    \begin{description}
        \item [Infomap:]
        This algorithm applies coding theory to compress streams representing the probability of paths in the network traversed by a random walker. The entropy of frequencies of each path is computed, and nodes are grouped when they are part of paths with less entropy in the coding compression~\cite{Rosvall2008}. 
        \item [Walktrap:]
        This algorithm computes the Euclidean distance of two communities based on a random walker's probability of being traversed. First, each node is the sole member of a community, and then the pair of communities with the lowest distance of the iteration merge~\cite{Pons2005}. It should be larger when nodes are in different communities and smaller for nodes in the same community.
        \item [Label-propagation:] This algorithm starts with a unique label for each node, and each node turns its label into the most common label in its neighbours. If there are ties, the label is chosen uniformly at random. This algorithm iterates until each node has the most common label in its neighbourhood~\cite{raghavan2007near}.
    \end{description}
\end{description}

\begin{figure}[ht!]
    \centering
    \includegraphics[width=0.95\textwidth]{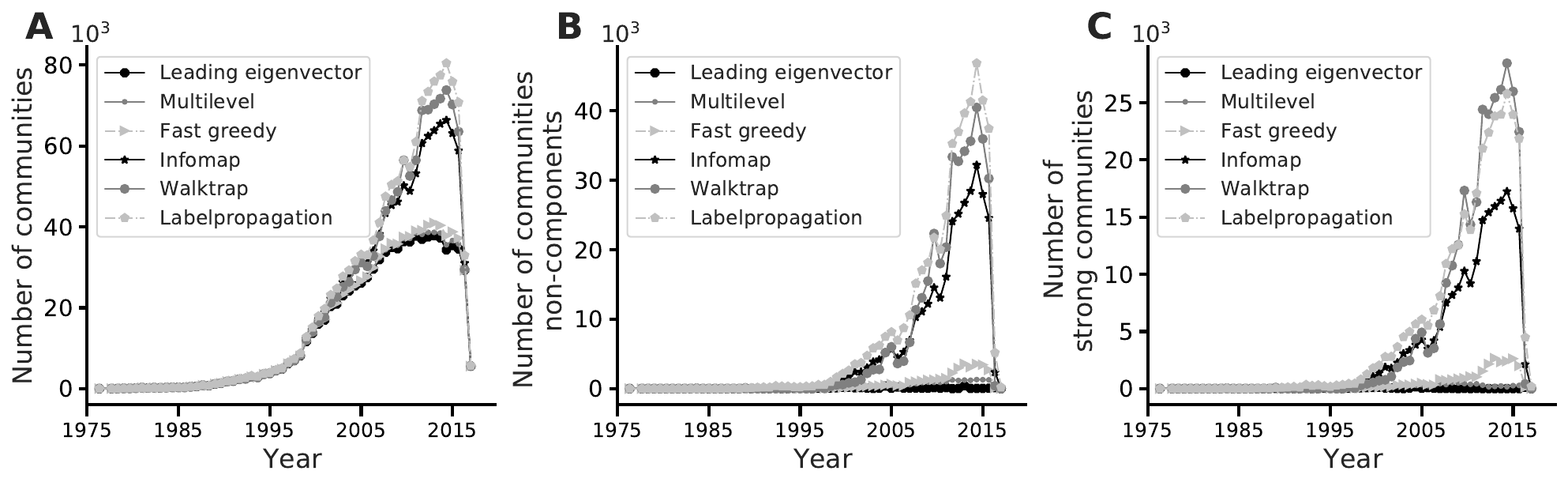}
    \caption{\textbf{Community detection results for the Semantic Scholar Strength network.} The left panel represents the number of communities over time, the central panel represents the number of communities that are not disconnected components, and the right panel is the number of strong communities based on the embeddedness of their nodes. A community is considered strong if all nodes in the community have $k_{i_C}/k_i>0.5$, where $k_{i_C}$ is the strength degree of the node $i$ inside the community $C$, and $k_i$ is the total strength degree of the node $i$ for a particular year.}
    \label{fig:Community_detection_analysis}
\end{figure}

After computing the community detection algorithms over 45 years, we can see how the number of communities grows over time with the size of the network. Those communities calculated with algorithms based on dynamics (Label-propagation, Walktrap and Infomap) have almost double the number of communities than the algorithms based on optimisation (Leading eigenvector, Multilevel, and Fast greedy) as we can see in \figurename~\ref{fig:Community_detection_analysis} in the left panel. The central panel shows the cause of the large difference in the number of communities: the algorithms based on dynamical processes have a larger number of connected communities, and we can conclude that dynamical algorithms tend to confound fewer communities with disconnected cliques.

We also analyse the communities' internal and external connectivity to select the appropriate algorithm. For the 6 algorithms, we study the strength of each community and its behaviour over time. For each community node, we calculate its embeddedness as its internal community degree strength $k_{i_C}$ over its total degree strength for the year. We label communities with all nodes embedded in the community as strong communities: $k_{i_C}/k_i>0.5$. Our results indicate that the community detection algorithms based on dynamic processes (Label-propagation, Walktrap and Infomap) have a larger number of strong communities than optimisation algorithms (Fast Greedy, Multilevel and Eigenvector), as we can see in \figurename~\ref{fig:Community_detection_analysis} in the right panel. From these results, we conclude that the detected communities are cohesive, and then the network presents a well-defined community structure over the 45 years timeline. In terms of connected and strong communities and their strength, the Label-propagation and Walktrap algorithms show better results than the other algorithms. For this analysis, we choose the results of communities from the Label-propagation algorithm to have a larger number of connected communities.

\section{Defining segregated communities}
\label{sec:segregated_communities_other_years}

For defining the segregated communities in 2006 and 2014, we followed the same procedure as in 2010 (\sectionname~4). We compute the 6 most used community detection algorithms and select the strongest communities based on the embeddedness of their nodes and the less confounded with connected components. As we can see in~\figurename~\ref{fig:Community_detection_analysis}, the results of the \textit{Label-propagation} algorithm show higher embeddedness in several years and less communities non-components in all years. 

\begin{figure}[ht!]
    \centering
    \includegraphics[width=0.8\textwidth]{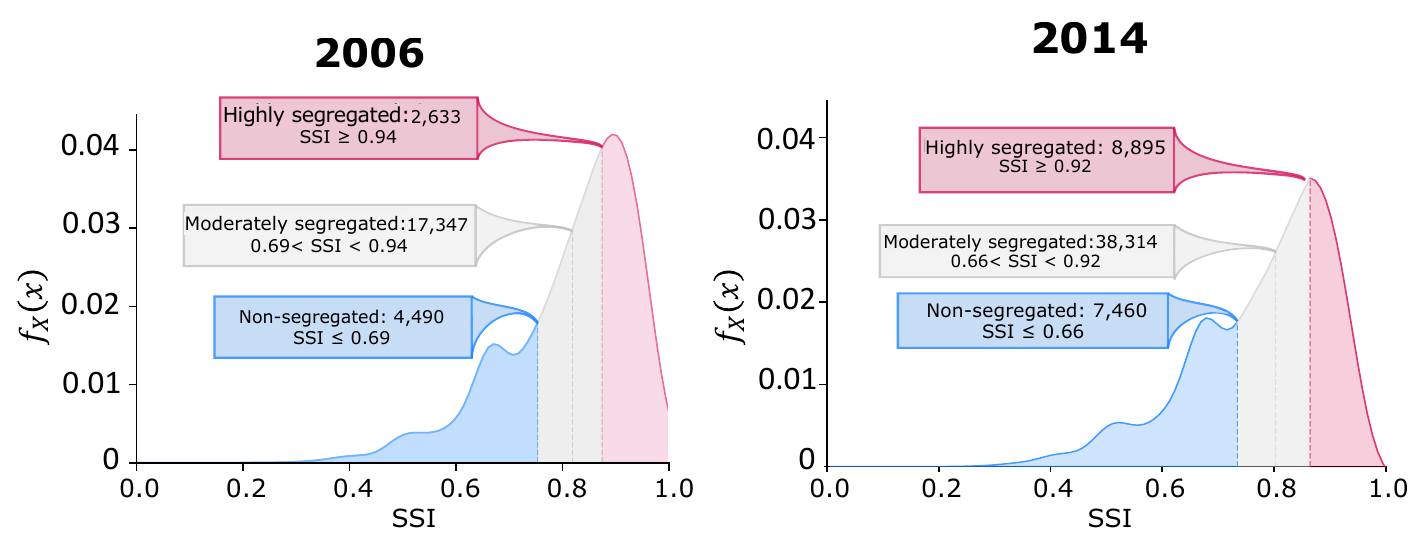}
    \caption{\textbf{Classifying communities as segregated and non-segregated for 2006 and 2014.} Probability density function (PDF) of the spectral segregation index (SSI) for 2006 and 2014. The plot is divided into 3 categories that denote non-segregated (light blue), moderately  (grey), and highly segregated (light red) communities. The complete procedure is in \sectionname~4.1. For the distribution, we use a Gaussian kernel density estimation with the ``rule of thumb'' for the bandwidth selection~\cite{scott2015multivariate}.}
    \label{fig:Partitions_SSI_2006_2014}
\end{figure}

Then, we compute the SSI for the resulting communities, and we divide the communities into 3 categories: non-segregated, moderately segregated, and highly segregated. 
Comparing the results in~\figurename~\ref{fig:Partitions_SSI_2006_2014} for 2006 and 2014, and ~\figurename~2
for 2010. We can see similar distributions in all years containing two peaks. In the 3 years, the largest category is moderately segregated, with 68\% of the communities for 2010, 71\% for 2006 and 70\% for 2014. The proportion of highly segregated communities increased from 11\% to 12\% from 2006 to 2010 and grew to 16\% in 2014. While the proportion of non-segregated communities had a low increment from 18\% to 19\%, while in 2014 it decreased to 14\%.

\section{Communities' core locations}
\label{sec:Network_communities_plot}
\begin{figure}[!h]
    \centering
    \includegraphics[width=0.7\textwidth]{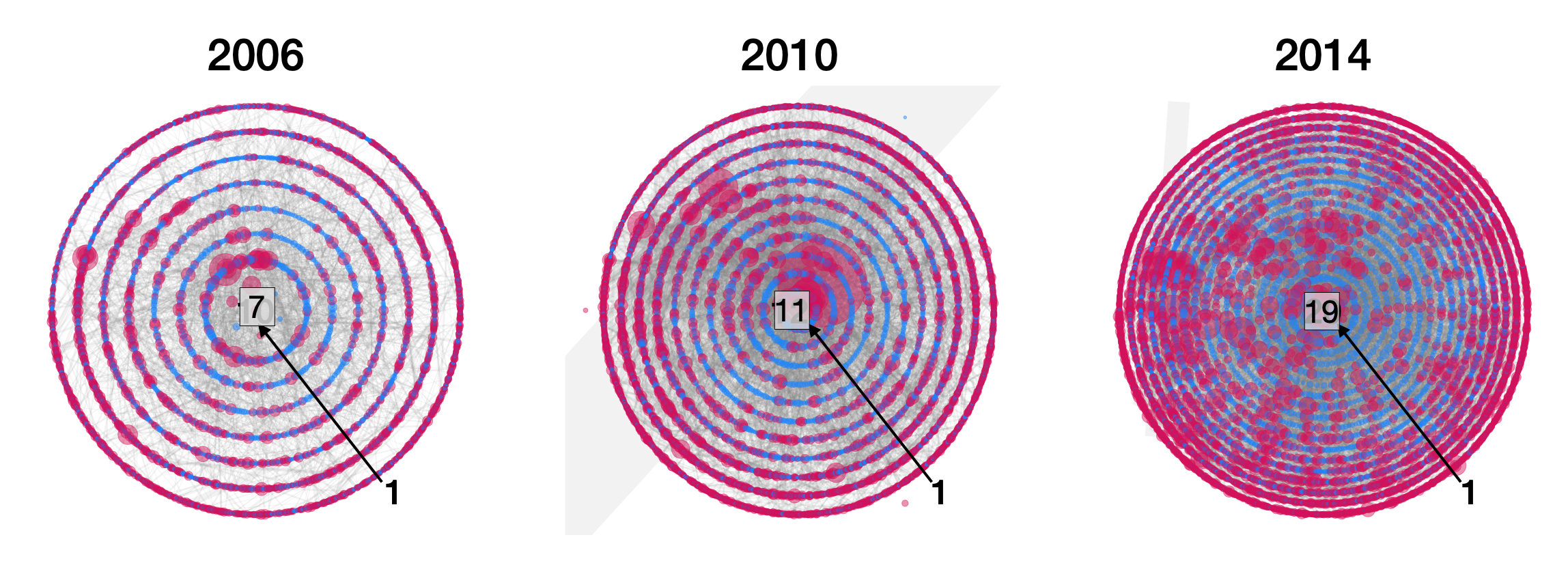}
    \caption{\textbf{Network of communities with the shell layout for highly segregated in red and non-segregated communities in blue for 2006, 2010, and 2014.} Each panel shows the results for one of the 3 years studied in this study. Each number refers to the k-core in which each community is located.}
    \label{fig:Shell_Layout_Communities_Network}
\end{figure}

This section deals with the network of communities, where each node is a community, and links are created if there are co-authorships among members of different communities. Once the networks are created, we compute the k-core decomposition~\cite{Batagelj2003}, and we locate each node (community) using a k-core layout in which the position corresponds to a specific core. In \figurename~\ref{fig:Shell_Layout_Communities_Network}, we show the communities network with nodes of non-segregated and highly segregated communities located in cores from the periphery (1) to the nucleus (7, 11, and 19 respectively for years 2006, 2010, and 2014). From the periphery to the network's nucleus, we can see how communities grow in size and that non-segregated (light blue) and highly segregated (light red) are in all  cores. 

As the network increases in size and connectivity (measured by the number of links and clustering coefficient), there are more cores, and patterns related to the difference in core location between non-segregated and highly segregated communities get clearer. As a first visual analyses, \figurename~\ref{fig:Shell_Layout_Communities_Network} shows how peripheral cores have more highly segregated communities (light red), and middle cores have more non-segregated communities (light blue). However, in the network's nucleus, because the size of the communities increases, it is difficult to see which is the segregation category of communities with the highest frequency.

However, in 2014, there is a clear pattern of more reddish peripheral cores while more blueish cores towards the nucleus. The figure does not show the moderately segregated communities, causing some cores to appear emptier. We analyse the statistical relationships among core position, size, and segregation categories in Sections~5
and ~\ref{sec:SSI_Core_Size_all_Years}.

\newpage
\section{Distributions of SSI for different community sizes}
\label{sec:PDFS_SSI_range_sizes}
\begin{figure}[ht!]  \includegraphics[width=0.8\textwidth,left]{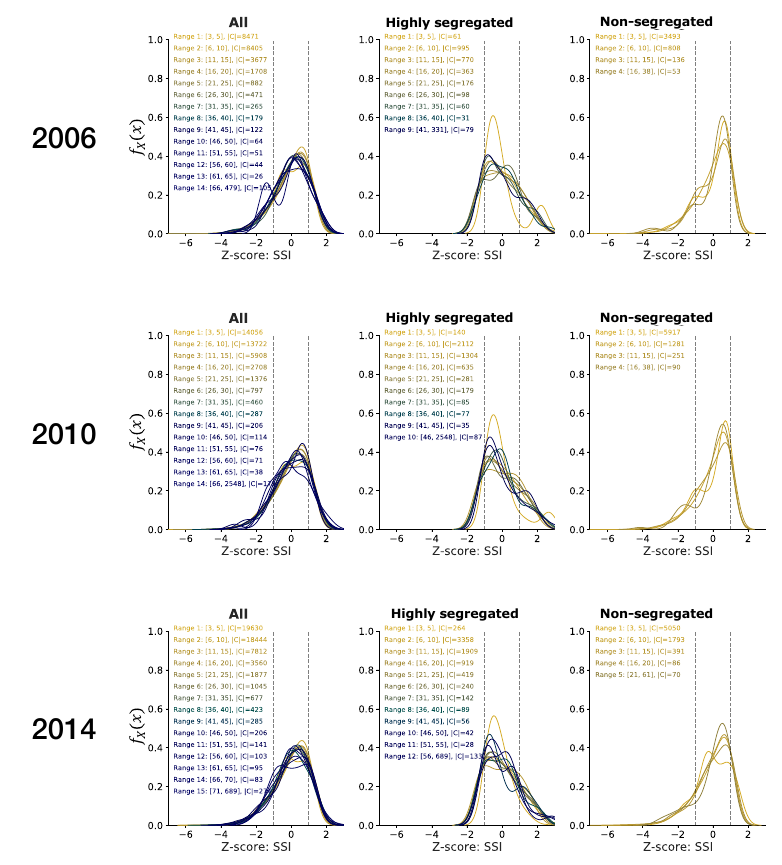}
    \caption{\textbf{SSI distribution for different range sizes when considering all, highly segregated and non-segregated communities.} Dashed lines represent a Z-Score of -1 and 1 as a visual guide for the communities with the behaviour expected by chance in their size range. Notice that the number of communities in the \textbf{Highly segregated} and \textbf{Non-segregated} panels range does not sum up the number of communities in the panel \textbf{All} because there were communities not classified in these two categories.}
    \label{fig:SSI_Distributions_without_dividing_by_size}
\end{figure}

This section compares the distribution of all highly segregated and non-segregated communities by different ranges of sizes, from 3-5 nodes in the first range to more than 66 nodes in the last range of 2006 and 2010, and more than 71 nodes for 2014, 
as shown in~\figurename~\ref{fig:SSI_Distributions_without_dividing_by_size}. As we can see in the second and third columns, some community sizes are not represented (as in the first column). We observe that highly segregated communities tend to be larger than non-segregated communities (the category all has moderately segregated communities as well). 

As we divided the communities into these categories using the PDF distribution of the SSI, which showed two peaks in Figures 2 and \ref{fig:Partitions_SSI_2006_2014}, the values of SSI for highly segregated communities tend to be 1 standard deviation lower than the mean value. In contrast, the values of SSI for non-segregated communities tend to be 1 standard deviation higher than their mean values. This result is interesting because if we separate the communities by their size, their SSIs are not following the trends of their group mean but tend to follow the mean values of the entire distribution: highly segregated communities represent the right portion of the distributions in the \textbf{All} panel, and non-segregated communities represent the left portion of that distribution. 

\newpage
\section{Structural metrics}
\label{sec:Comparing_Structural_metrics}
In this section, we compare the probability density functions (PDFs) of size, density, clustering, and core position for 2006, 2010 and 2014. In \sectionname~\ref{sec:PDFs_all_sizes}, we compare the metrics without dividing the communities by size ranges for the 3 years, while in the \sectionname~\ref{sec:Metrics_Null_models}, we compute compare the results of the three years with null models correcting by size.

\subsection{Highly segregated and non-segregated communities without differentiating by size}
\label{sec:PDFs_all_sizes}
\begin{figure}[!h]
    \centering
    \includegraphics[width=0.95\textwidth]{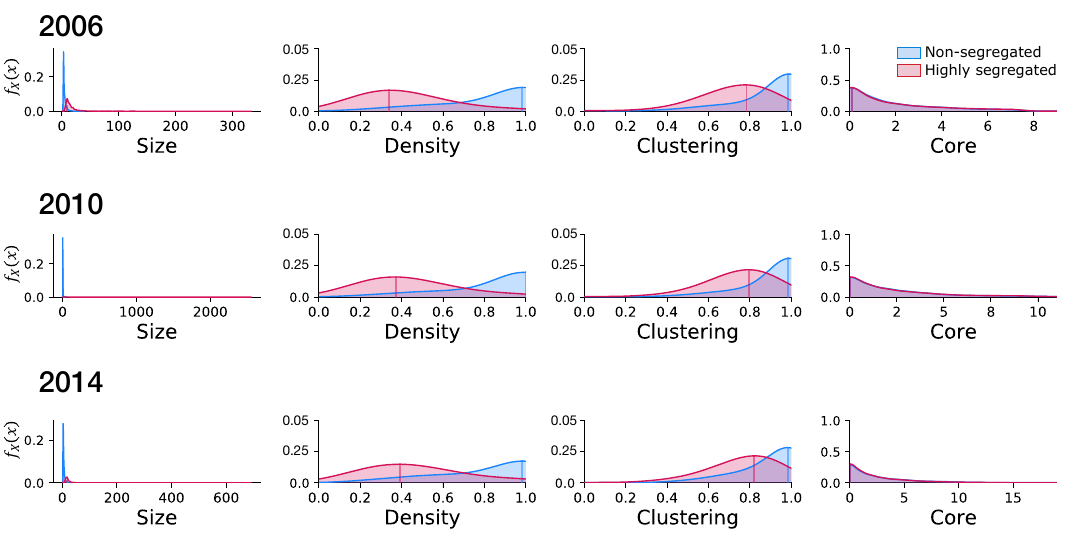}
    \caption{\textbf{Distribution of topological and core values of non-segregated and highly segregated communities without dividing by small or large communities.} Panels represent the probability density functions (PDF) for the size, density, clustering and core position of highly segregated (red) and non-segregated (blue) communities for 2006, 2010, and 2014.}
    \label{fig:Distributions_without_dividing_by_size}
\end{figure}

This section shows the PDFs of four structural metrics for highly segregated (light red) and non-segregated (light blue) communities without separating the communities by size range for 2006, 2010, and 2014. From~\figurename~\ref{fig:Distributions_without_dividing_by_size}, we observe that highly segregated communities tend to be larger, less dense, and less clustered, and there are no differences in the core position of the communities. Interestingly, the distribution of density and clustering coefficient does not have apparent changes over the years, with a peak of density around 0.4 for highly segregated and 1 for non-segregated communities. Moreover, for the clustering coefficient, there is a peak of around 0.8 for highly segregated communities and around 1 for non-segregated communities. Those density and clustering coefficient values can be driven mainly by non-segregated communities being smaller than highly segregated communities, as we see in the first column for the 3 years. In addition, there tend to be more communities of both segregation categories in smaller cores (towards the periphery). In contrast, the number of cores increases with the years and the patterns we saw in \figurename~\ref{fig:Shell_Layout_Communities_Network} disappear if we compare all the communities. 

The results  of this subsection are misleading, as we see in Section~5,
the results change when we divide the communities by size because the size of the communities mainly drives the density and clustering coefficient. Still, the core will be analysed in the following sections, and we argue that comparing the analysis with and without dividing by categories makes the main manuscript more reliable.

\subsection{Community metrics comparison with Z-Scores}
\label{sec:Metrics_Null_models}
Therefore, this part of the analysis aims to measure the significant differences between non-segregated and highly segregated communities in the same size range. Because size is the control metric, we solely analyse the density, clustering coefficient, and core position. We perform Z-Scores comparisons for each metric across segregation categories controlling for size. For example, we take the density of each highly segregated community in the size range: [3,5], and we calculate a Z-Score with the density of all non-segregated communities in the same size range. Then, we analyse the PDF distributions of all the Z-Scores as shown in Figures~\ref{fig:Topological_metrics_null_models_oposite_category_divided_2006},~\ref{fig:Topological_metrics_null_models_oposite_category_divided_2010}, and~\ref{fig:Topological_metrics_null_models_oposite_category_divided_2014} for 2006, 2010, and 2014, respectively. We also compare the PDFs without separating them by size, and the results and analyses are in Section~\ref{sec:Null_models_all_sizes}.

In line with the results of the previous section, the Z-Scores comparisons show no significant differences in the density and clustering coefficient for both types when we control for size. The Z-Scores have their largest peak at zero for most cases, meaning that these communities are not significantly different from other communities of the same range size with a different segregation category. There is a difference in the density and clustering coefficient of small communities of size [3,5] in which highly segregated communities show to be less dense and less clustered (Figures~\ref{fig:Topological_metrics_null_models_oposite_category_divided_2006},~\ref{fig:Topological_metrics_null_models_oposite_category_divided_2010}, and~\ref{fig:Topological_metrics_null_models_oposite_category_divided_2014} first row, first and second panels). The main difference is consistent with the core position of the communities. When controlling for size, all communities of different sizes have similar patterns: highly segregated communities are towards the periphery compared with the non-segregated ones that are towards the nucleus in all range sizes of comparison (third column of the previous figures).

\subsubsection{Highly segregated and non-segregated communities Z-Scores without differentiating by size}
\label{sec:Null_models_all_sizes}
\begin{figure}[!h]
    \centering
    \includegraphics[width=0.80\textwidth]{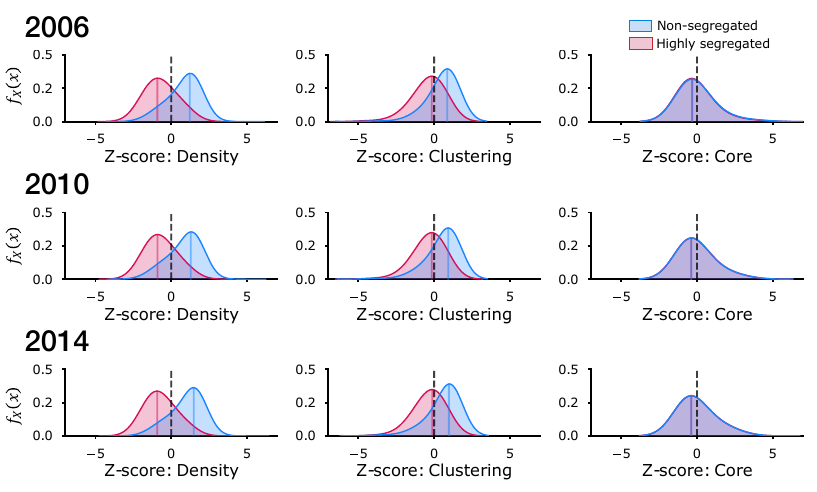}
    \caption{\textbf{Comparison of the topological and core position of highly segregated and non-segregated communities with communities of the same range size without differentiating by size} Panels represent the probability density functions (PDF) for the Z-Score of comparing the density, clustering and core position of highly segregated (red) and non-segregated (blue) communities with opposite communities, i.e. highly segregated compared with non-segregated of the same size. The PDFs were computed using just the Z-Scores of comparisons that had at least 30 communities of the opposite category and the same size to compare. The dashed line in zero represents no significant difference, above zero means a higher variable value, and below zero implies smaller values.}
    \label{fig:Topological_metrics_null_models_oposite_category_non_divided}
\end{figure}

This section shows the results from the Z-Score analyses by comparing the topological metrics and core position of highly segregated and non-segregated communities without differentiating them by size. Each community is compared with at least 30 communities of the opposite category. From ~\figurename~\ref{fig:Topological_metrics_null_models_oposite_category_non_divided}, we can observe that segregated communities tend to be less dense and less clustered, and there are no differences in the core position of the communities. However, as we see in Section~5,
the results change when we differentiate the communities by size because the density and clustering coefficient are not different between both segregation categories. Still, the core position has some differences for larger communities. Here, we argue that the analysis of the Main Manuscript is more accurate because the community's size mainly drives density and clustering values.

\newpage
\newpage
\subsubsection{Highly segregated and non-segregated communities Z-Scores differentiating by size}
\begin{figure}[!h]
    \includegraphics[width=0.80\textwidth, left]{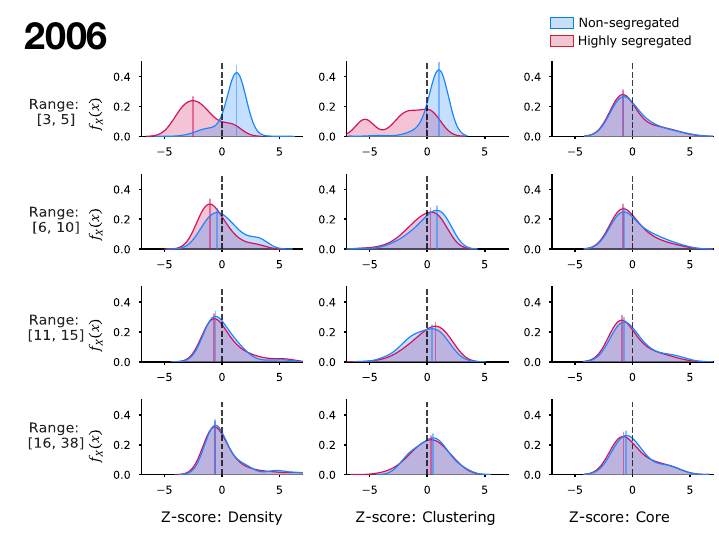}
    \caption{\textbf{Comparison of the topological and core position of segregated and non-segregated communities with communities of the same range size differentiating by size for 2006} Panels represent the probability density functions (PDF) for the Z-Score of comparing the density, clustering and core position of highly segregated(red) and non-segregated(blue) communities with opposite communities, i.e. highly segregated compared with non-segregated of the same size. The PDFs were computed using just the Z-Scores of comparisons that had at least 30 communities of the opposite category and the same size to compare. The dashed line in zero represents no significant difference, above zero represents a higher variable value, and below zero implies smaller values. }
    \label{fig:Topological_metrics_null_models_oposite_category_divided_2006}
\end{figure}

\begin{figure}[!h]
    \includegraphics[width=0.80\textwidth, left]{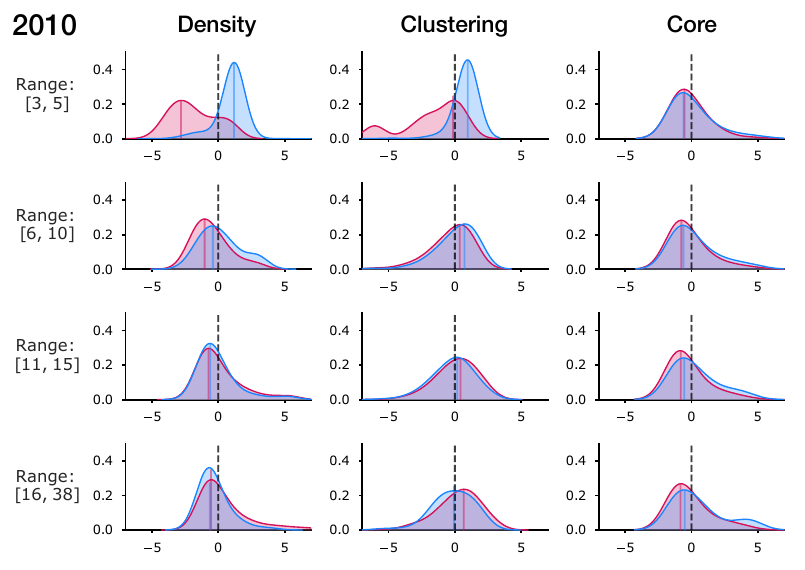}
    \caption{\textbf{Comparison of the topological and core position of segregated and non-segregated communities with communities of the same range size differentiating by size for 2010} Panels represent the probability density functions (PDF) for the Z-Score of comparing the density, clustering and core position of segregated (red) and non-segregated (blue) communities with opposite communities, i.e. highly segregated compared with non-segregated of the same size. The PDFs were computed using just the Z-Scores of comparisons with at least 30 communities of the opposite category and the same size to compare. The dashed line in zero represents no significant difference, above zero means a higher variable value, and below zero implies smaller values. }
    \label{fig:Topological_metrics_null_models_oposite_category_divided_2010}
\end{figure}

\begin{figure}[!h]
\includegraphics[width=0.80\textwidth, left]{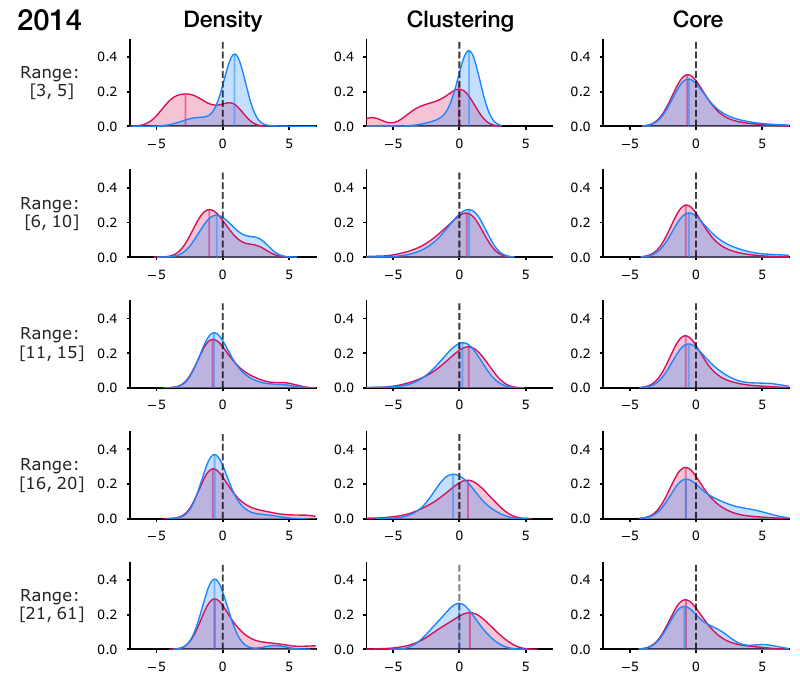}
    \caption{\textbf{Comparison of the topological and core position of segregated and non-segregated communities with communities of the same range size differentiating by size for 2014} Panels represent the probability density functions (PDF) for the Z-Score of comparing the density, clustering and core position of highly segregated (red) and non-segregated (blue) communities with opposite communities, i.e. highly segregated compared with non-segregated of the same size. The PDFs were computed using just the Z-Scores of comparisons that had at least 30 communities of the opposite category and the same size to compare. The dashed line in zero represents no significant difference, above zero means a higher variable value, and below zero implies smaller values. }
    \label{fig:Topological_metrics_null_models_oposite_category_divided_2014}
\end{figure}

This section shows the results from the Z-Score analyses by comparing the topological metrics and core position of highly segregated and non-segregated communities with communities of the same size range differentiating them by size. Each community is compared with at least 30 communities of the opposite category. From ~\figurename~\ref{fig:Topological_metrics_null_models_oposite_category_divided_2006}, we can observe that for 2006 there are some differences when communities are small (3 to 5 nodes) in which highly segregated communities tend to be less dense and clustered than non-segregated communities but there are no differences by core or in any of the metrics when the communities are in higher range sizes. However, when we see the results for 2010 and 2014 in Figures~\ref{fig:Topological_metrics_null_models_oposite_category_divided_2010} and \ref{fig:Topological_metrics_null_models_oposite_category_divided_2014} we see clear differences in the  core position of all range sizes where highly segregated communities are in lower cores than non-segregated communities. In addition, for small communities there are no much differences in density and clustering but when the size range increase to more than 10 nodes. highly segregated communities seem denser and more clustered than non-segregated communities. We infer this differences in 2010 and 2014 are due to the growth of the network.

\newpage
\newpage
\subsubsection{Size and segregation of communities in different cores}
\label{sec:SSI_Core_Size_all_Years}
\begin{figure}[ht!]
    \centering
    \includegraphics[width=0.95\textwidth]{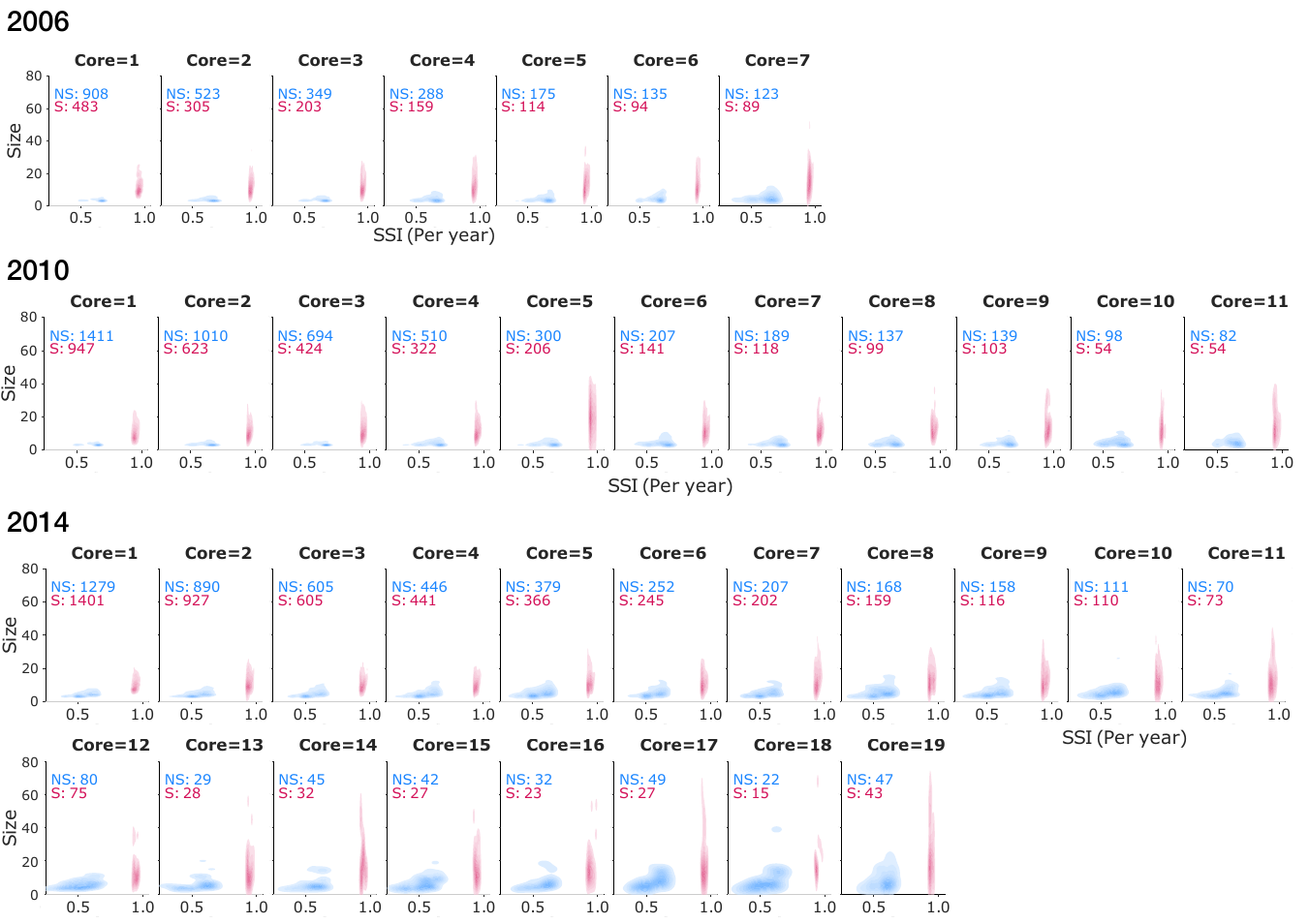}
    \caption{\textbf{Relation between communities size, segregation and core position in the network for 2006, 2010 and 2014} Each panel represents the Kernel Density Estimation (KDE) for size in the y-axis, SSI in the x-axis and the core position of the communities. In red are those highly segregated, while non-segregated communities are in blue. Darker colours show a higher proportion of communities, while lighter colours represent fewer ones. Each panel shows the number of communities of each type used to compute the corresponding plot. The first cores show communities in the periphery, while the 6th and 13th cores show the communities in the network's nucleus.}
    \label{fig:Size_SSI_Core_2006_2014}
\end{figure}

We aim to understand the relationship among the variables that have shown higher differences between non-segregated and highly segregated communities: size and core position. To this end, we use kernel density estimators (KDE) to compare the 3 variables simultaneously, using a fair representation of their probability density functions. In \figurename~\ref{fig:Size_SSI_Core_2006_2014}, we show the KDE results with smooth 2D curves for non-segregated and highly segregated to compare the 3 variables with more pronounced differences: SSI, size, and core position. The results remain similar over the 3 years. The behaviour of highly segregated communities being larger remains in the 3 years. Also, when we go towards the network's nucleus, increasing the core position, the size of the communities gets larger, and this behaviour happens more for highly segregated communities, as we can see in~\figurename~\ref{fig:Size_SSI_Core_2006_2014} for 2006, 2010 and 2014.

When comparing the number of communities per core, there are
more small communities towards the periphery and fewer but larger
communities towards the nucleus, as we can see in
\tablename~\ref{tab:com_size_core}.
When the core position of the communities increases (towards the nucleus), 
the total number of researchers in those communities decreases, with fewer researchers in the nucleus of the
network (\tablename~\ref{tab:com_size_core}). Then, there are smaller communities 
in the periphery but with a larger number of researchers. This finding aligns with
previous results~\cite{Malvestio2020}, where the k-core structure of
some empirical and randomised networks were shown to be explained by their
community structure. 

\begin{table}[ht!]
\caption{number of communities ({\em NC}), Number of researchers ({\em NR}), number of researchers per community ({\em NR$/$NC}), number of researchers in non-segregated communities ({\em NR$_{\text{NS}}$}) per core, and number of researchers in highly segregated communities ({\em NR$_{\text{HS}}$}). }
\resizebox{1\textwidth}{!}{%
      \begin{tabular}{ l r r r r r r r r r r r r r }
        \toprule
        Core & 1 & 2 & 3 & 4 & 5 & 6 & 7 & 8 & 9 & 10 & 11\\
        \midrule
        {\em NC} & 14,542 & 7,445 & 5,165 & 3,577 & 2,681 & 1,642 & 1,193 & 1,036 & 789 & 864 & 539 \\
        {\em NR} & 133,416 & 73,748 &	52,016 & 37,222 & 29,987 & 21,285 & 13,426 & 11,556 & 9,085 &	10,708 &  6,961 \\
        {\em NR/NC} & 9.17 & 9.91 & 10.07 & 10.41 & 11.19 & 12.96 & 11.25 & 11.15 & 11.51 & 12.39 & 12.91 \\
        {\em NR$_{\text{HS}}$} & 24,056 &	13,975 & 8,806 & 6,510 & 4,786 &	6,164 & 2,222 & 1,812 & 1,607 &	1,774 & 1,168 \\
        {\em NR$_{\text{NS}}$} & 12,047 &	6,411 &	4,601 & 3,170 & 2,490 &	1,437 & 1,104 & 927 & 699 & 713 & 549 \\

        \bottomrule
      \end{tabular}
      }
      \label{tab:com_size_core}
\end{table}

\newpage
\newpage
\section{Citations of segregated and non-segregated communities}
\label{sec:Citations_communities}
\begin{figure}[ht!]
    \centering
    \includegraphics[width=0.80\textwidth]{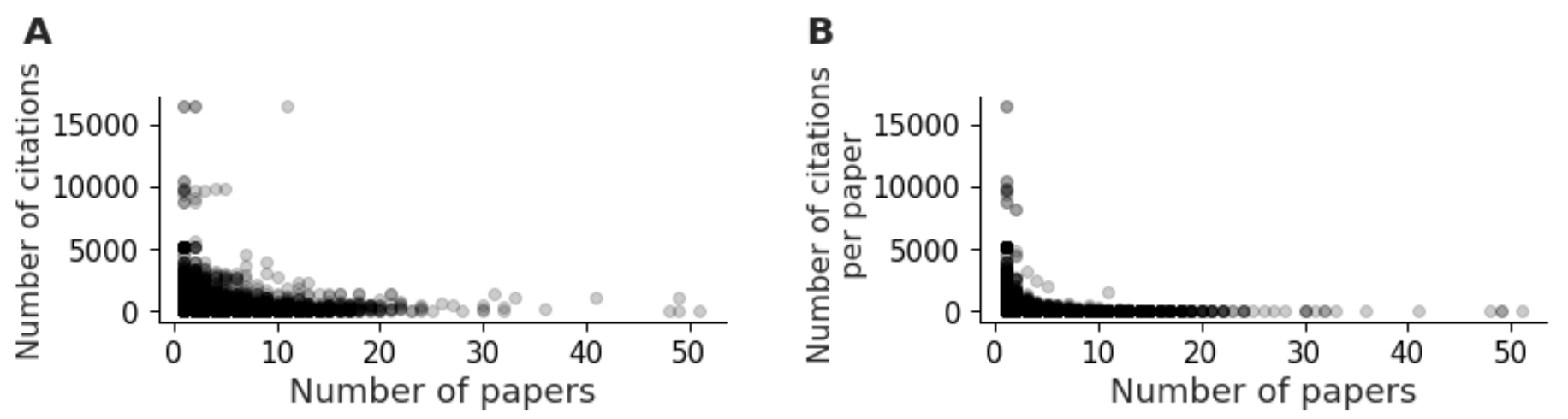}
    \caption{\textbf{Number of papers vs the number of citations and citations per paper in 2010} Scatter plot for the number of papers each researcher published in 2010 and the number of citations and citations per paper until 2020.}
    \label{fig:Production_vs_citations}
\end{figure}

This section shows the results and procedure to compute and compare researchers' citations in non-segregated and highly segregated communities for 2006, 2010, and 2014. For our main year of analysis, 2010, we compare the number of papers published with the number of citations and citations per paper received until 2020. In previous literature, the number of citations an author receives has been related to their number of publications~\cite{Huang2020}. Still, our results show a non-linear relationship between the number of published papers in one year and the number of citations gained by those papers after ten years of publication (the publication year was in 2010, and the citations count is until 2020), as shown in \figurename~\ref{fig:Production_vs_citations}. With a low Spearman correlation of 0.29 for the number of citations and 0.09 for the number of citations per paper with $p$-values smaller than 0.05 in both cases, we see a heavy tail decay in the plots. Then, a researcher with a large number of publications do not necessarily imply they have a large number of citations. 

\begin{figure}[ht!]
    \centering
    \includegraphics[width=0.95\textwidth]{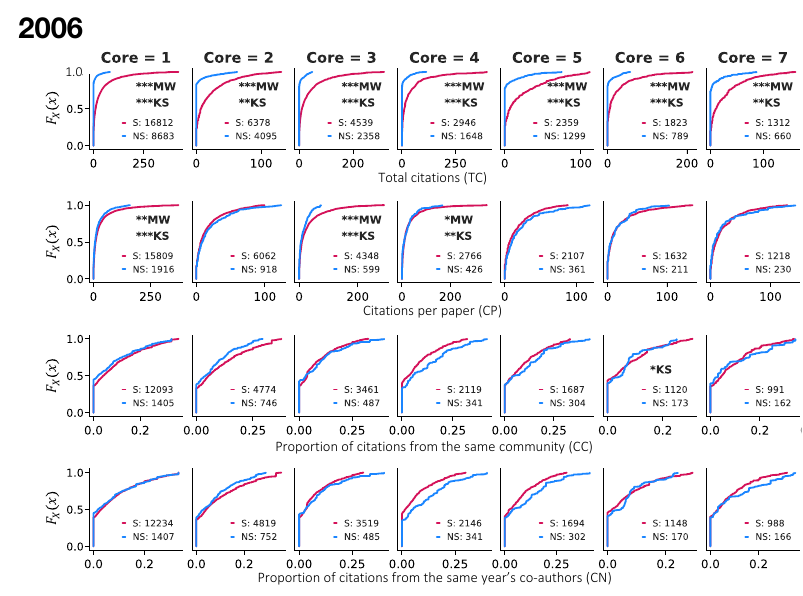}
    \caption{\textbf{Citation metrics for researchers in communities of different segregation categories and core positions for 2006.} Each row represents the cumulative density function (CDF) for the total citations (TC), the citations per paper (CP), the proportion of citations from the same community (CC), and the proportion of citations from the same year's co-authors (CN). The code of colours is: light red for highly segregated (S) and blue for non-segregated communities (NS). Letters \textbf{KS} or \textbf{MW} appear when there are significant $p$-values values for Kolmogorov-Smirnov (different distribution shapes) and Mann-Whitney (different distribution medians) for the CDFs of non-segregated and highly segregated communities. Significance levels are denoted as follows: * $<$ 0.1, ** $<$ 0.05, and *** $<$ 0.01.}
    \label{fig:Citations_plot_allcores_2006}
\end{figure}

\begin{figure}[ht!]
    \centering
    \includegraphics[width=0.95\textwidth]{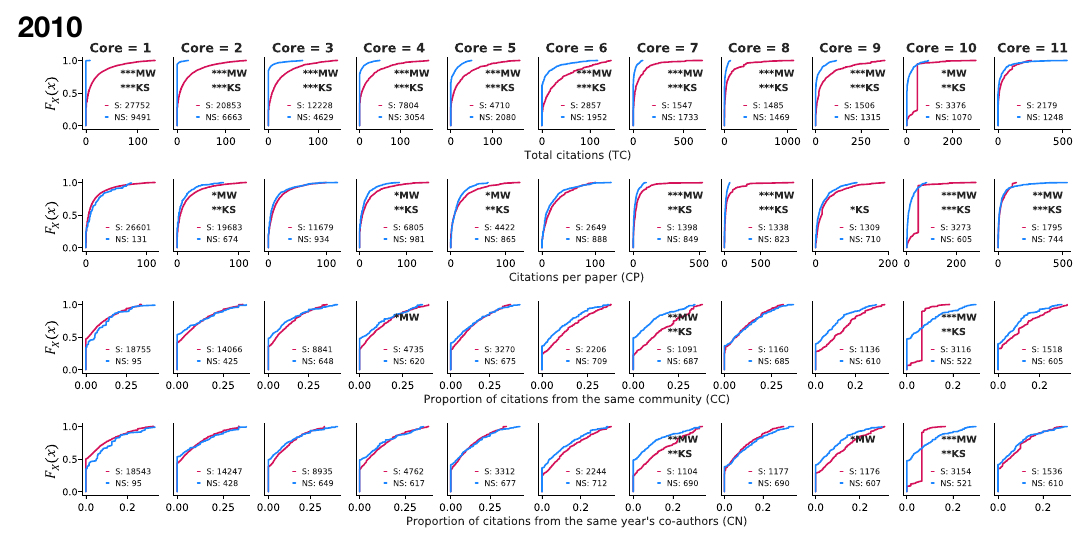}
    \caption{\textbf{Citation metrics for researchers in communities of different segregation categories and core positions for 2010.} Each row represents the cumulative density function (CDF) for the total citations (TC), the citations per paper (CP), the proportion of citations from the same community (CC), and the proportion of citations from the same year's co-authors (CN). The code of colours is: light red for highly segregated (S) and blue for non-segregated communities (NS). Letters \textbf{KS} or \textbf{MW} appear when there are significant $p$-values values for Kolmogorov-Smirnov (different distribution shapes) and Mann-Whitney (different distribution medians) for the CDFs of non-segregated and highly segregated communities. Significance levels are denoted as follows: * $<$ 0.1, ** $<$ 0.05, and *** $<$ 0.01.}
    \label{fig:Citations_plot_allcores_2010}
\end{figure}

\begin{figure}[ht!]
    \centering
    \includegraphics[width=0.95\textwidth]{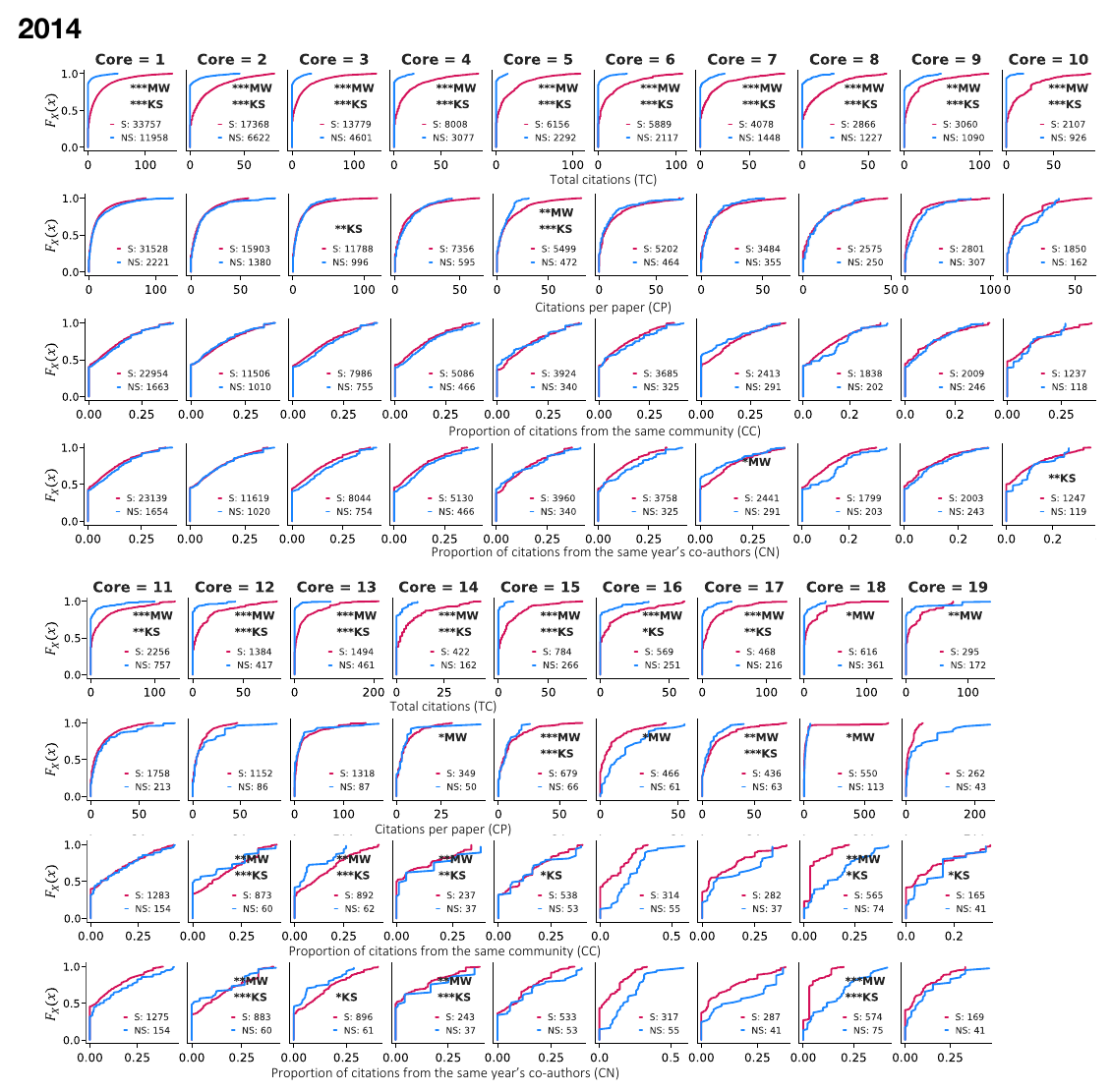}
    \caption{\textbf{Citation metrics for researchers in communities of different segregation categories and core positions for 2014.} Each row represents the cumulative density function (CDF) for the total citations (TC), the citations per paper (CP), the proportion of citations from the same community (CC), and the proportion of citations from the same year's co-authors (CN). The code of colours is: light red for highly segregated (S) and blue for non-segregated communities (NS). Letters \textbf{KS} or \textbf{MW} appear when there are significant $p$-values values for Kolmogorov-Smirnov (different distribution shapes) and Mann-Whitney (different distribution medians) for the CDFs of non-segregated and highly segregated communities. Significance levels are denoted as follows: * $<$ 0.1, ** $<$ 0.05, and *** $<$ 0.01.}
    \label{fig:Citations_plot_allcores_2014}
\end{figure}

Then, we analyse if there are differences in the citation patterns depending on the segregation category of the communities. In~\figurename~\ref{fig:Citations_plot_allcores_2006},~\ref{fig:Citations_plot_allcores_2010}, and ~\ref{fig:Citations_plot_allcores_2014}, we show the CDFs for the four metrics to analyse the research impact: for understanding the citation numbers we have: {\em (i)} total number of citations and {\em (ii)} citations per paper; and for understanding the citation sources we have: {\em (iii)} proportion of citations from the same community and {\em (iv)} proportion of all citations from the same year's co-authors. 
We use two statistical tests to compare the CDFs of non-segregated and highly segregated communities: Kolmogorov-Smirnov (KS) and Mann-Whitney (MW). The first test compares the shape of the distributions, and the second compares the differences between medians.

First, we investigate whether the number of internal papers correlates with {\em (i)} the total number of citations and {\em (ii)} the average number of citations per paper. In this section we analyse the first two rows of~Figures~\ref{fig:Citations_plot_allcores_2006},~\ref{fig:Citations_plot_allcores_2010}, and ~\ref{fig:Citations_plot_allcores_2014} for 2006, 2010, and 2014, respectively. The results of the 3 years are consistent with researchers in highly segregated communities (red-longer tails) having more total citations than researchers in non-segregated communities (blue-shorter tails) (First row). However, when correcting the citations by the number of papers, the differences among researchers in different segregation categories are less significant. Some cores near the nucleus have researchers in non-segregated communities with higher citations per paper (second row).

Then, we analyse the CDFs for {\em (iii)} the proportion of citations from the same community (CC) and {\em (iv)} the proportion of citations from the same year’s co-authors (CN). In this section we analyse the last two rows of~\figurename s~\ref{fig:Citations_plot_allcores_2006},~\ref{fig:Citations_plot_allcores_2010}, and ~\ref{fig:Citations_plot_allcores_2014} for 2006, 2010, and 2014. The results of the 3 years are consistent. For cores towards the periphery, there are no significant differences in the proportion of citations received by community members (third row) or co-authors in the same year (fourth row). In middle cores, there are some cases in which researchers in non-segregated communities have a larger proportion of those metrics than researchers in highly segregated communities. However, in the 2010 and 2014 networks' nucleus, the researchers in highly segregated communities have more internal citations (\figurename s~\ref{fig:Citations_plot_allcores_2010} and \ref{fig:Citations_plot_allcores_2014}).

\newpage
\newpage
\section{Using \textit{Infomap} communities}
\label{sec:Infomap_analysis}

\subsection{Defining segregated communities}
\begin{figure}[ht!]
    \centering
    \includegraphics[width=0.5\textwidth]{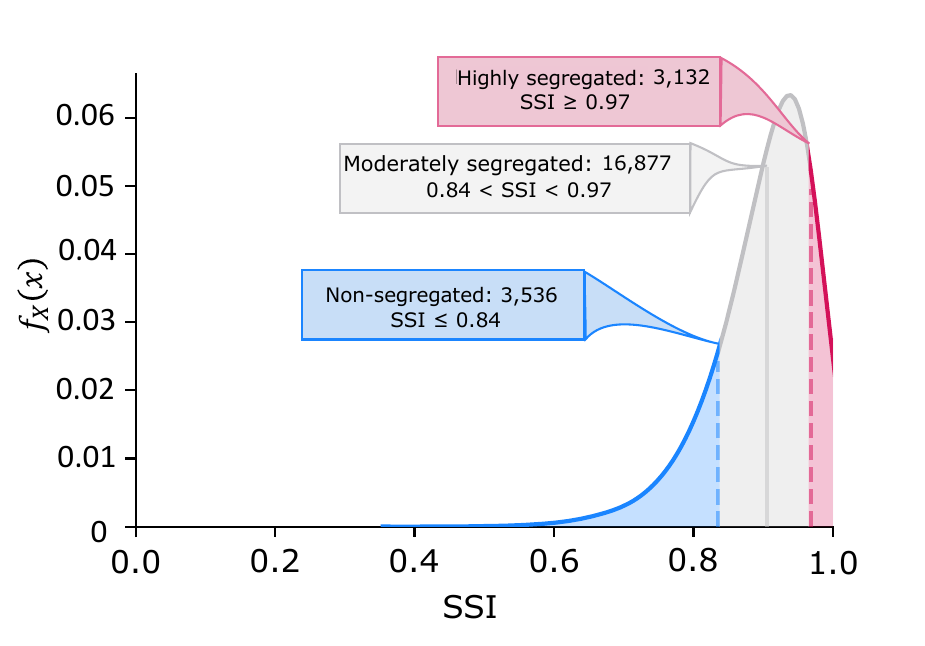}\caption{\textbf{Classifying communities as highly segregated and non-segregated \textit{Infomap} 2010} Probability density function (PDF) of the spectral segregation index (SSI) for Infomap communities in 2010. The plot is divided into 3 categories that denote non-segregated (highly segregated) communities with a value of SSI smaller(larger) than 1 standard deviation from the mean value of the SSI distribution. In grey are those communities within 1 standard deviation categorised as moderately segregated communities and are not part of the following analysis.}
    \label{fig:Partitions_SSI_Infomap_2010}
\end{figure}

This section shows the results of our analyses when studying a community partition to the co-authorship network of 2010 with the \textit{Infomap} algorithm. In total, we studied 23,545 communities. We compute the SSI for the resulting communities following the procedure described in \sectionname~4. We divide the communities into 3 categories: non-segregated, moderately segregated, and highly segregated. Comparing~\figurename~\ref{fig:Partitions_SSI_Infomap_2010}. There is a similar pattern in the PDF of the SSI for both algorithms but a more skewed distribution towards larger SSI values for communities found with \textit{Infomap}.

\begin{figure}[!h]
    \centering
    \includegraphics[width=0.7\textwidth]{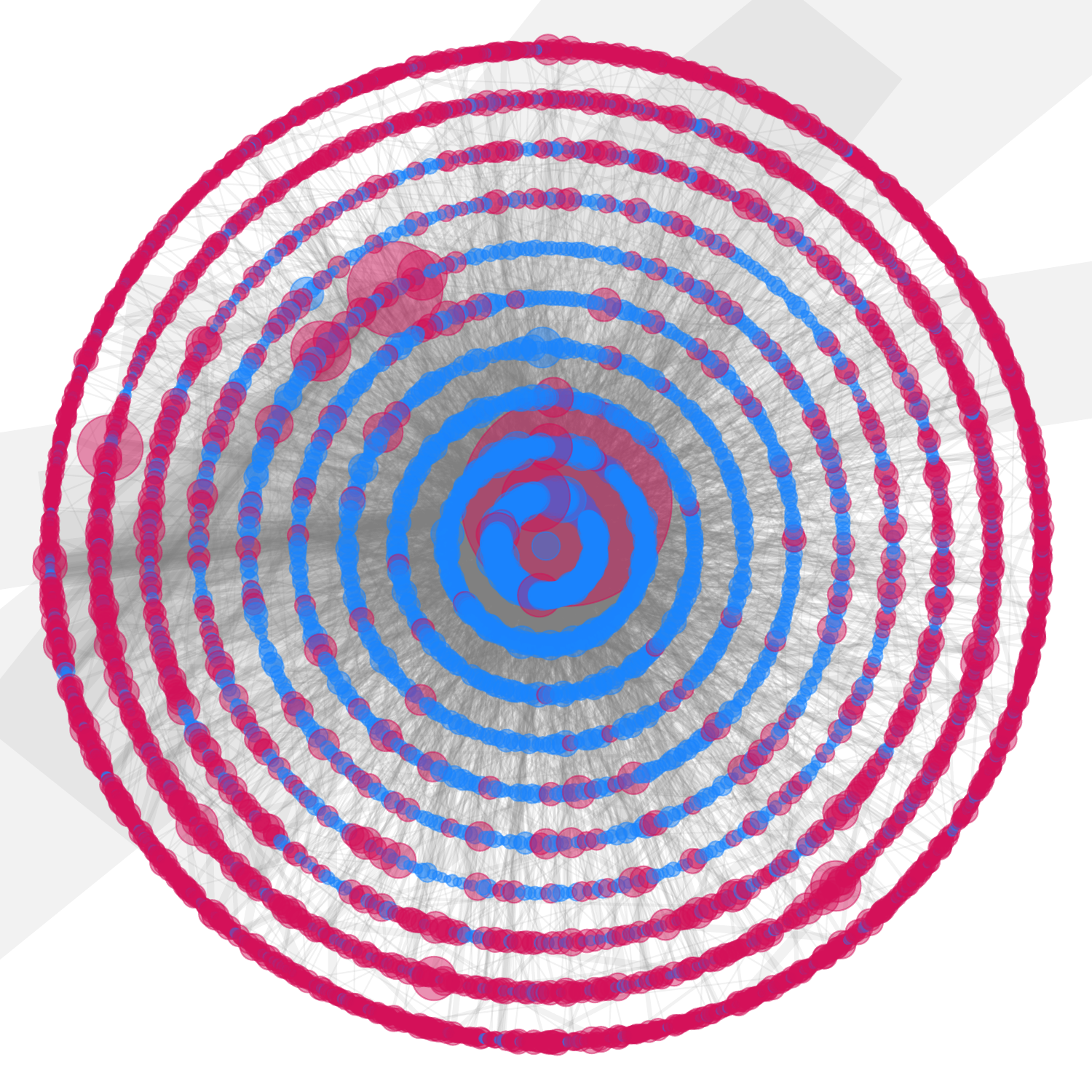}
    \caption{\textbf{Network of communities of \textbf{Infomap} with the shell layout for highly segregated in red and non-segregated communities in blue for 2010.} Each panel shows the results for one of the 3 years studied in this study. Each number refers to the k-core in which each community is located.}
    \label{fig:Shell_Layout_Communities_Network_Infomap}
\end{figure}

This section shows the network of communities with each community located in a core from the periphery (1) to the nucleus~(11) for 2010 with \textit{Infomap} (\figurename~\ref{fig:Shell_Layout_Communities_Network_Infomap}). From the periphery to the network's nucleus, we can see how communities grow in size and that non-segregated (blue) and highly segregated (red) are in all network cores. However, periphery cores have more highly segregated communities, and middle cores have more non-segregated communities. Because the size of highly segregated communities increases in the network's nucleus, it is difficult to see which communities are the most dominant. We analyse the statistical relationships among core position, size, and segregation categories in \sectionname~\ref{sec:structural_metrics_infomap}.

\subsection{Structural metrics}
\label{sec:structural_metrics_infomap}
We analyse the PDFs for size, density, clustering and core position of the \textit{Infomap} communities. We find no difference in size or density, but when communities are larger, highly segregated ones are more clustered. Compared with the results of \textit{Label-propagation}, both algorithms have similar results when comparing the PDFs of size, density, clustering coefficient, and core position between non-segregated and highly segregated communities, demonstrating that our results are not dependent of one community detection algorithm. When the community's size increases: {\em i)} highly segregated communities have higher density and clustering and stay in peripheral cores, and {\em ii)} non-segregated communities remain with similar values of density and clustering but locate in cores nearer the nucleus. \textit{Label-propagation} always has more highly segregated researchers than non-segregated, while for \textit{Infomap}, it changes: there are more highly segregated researchers in the periphery and more non-segregated researchers in the nucleus.

\begin{figure}[p]
    \centering
    \includegraphics[width=0.98\textwidth]{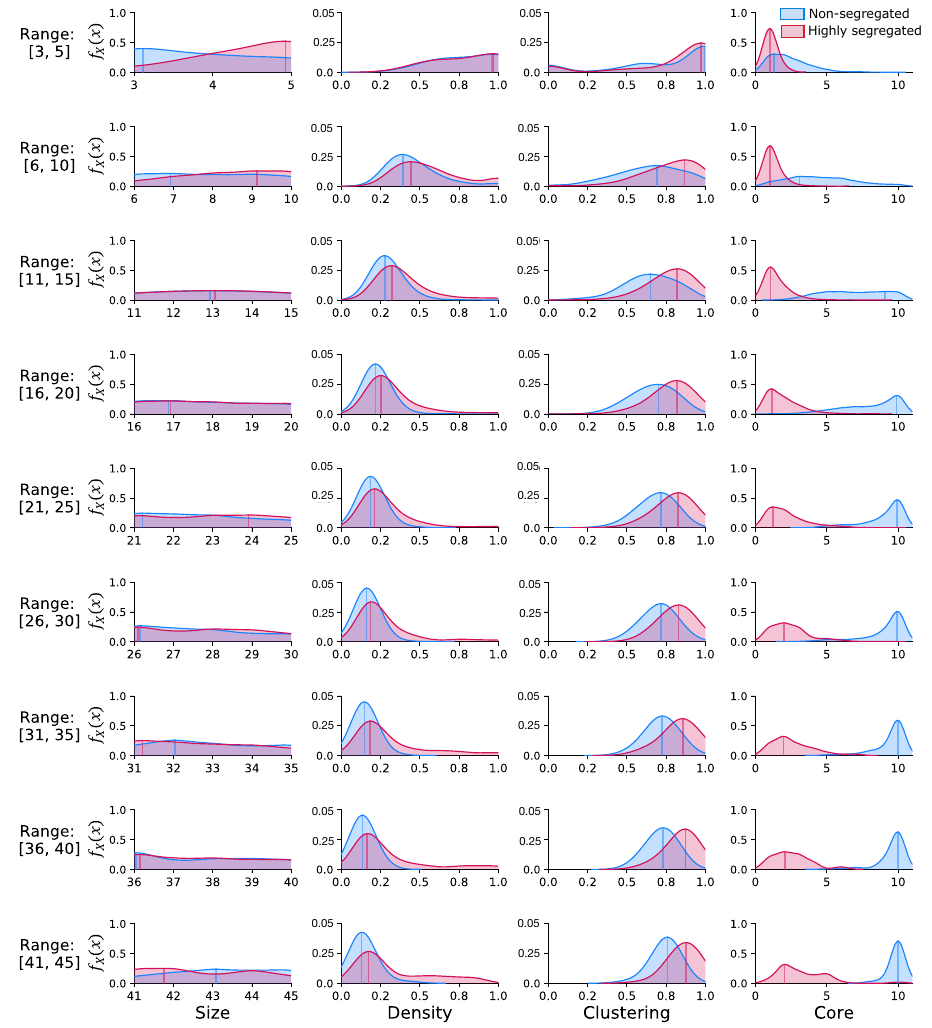}
    \caption{\textbf{Topological and core position differences among segregated and non-segregated communities for 2010 with \textbf{Infomap}} The panels represent the probability density functions (PDF) in each column for the size, density, clustering, and core position of highly segregated(red) and non-segregated(blue). Each row represents communities with the number of researchers written in the Range label. The PDFs were computed using just the communities in the sample after separating by size, hence the different y-axis limits.}
    \label{fig:Topological_metrics_2010_Infomap}
\end{figure}

\subsection{Citations of highly segregated and non-segregated communities}
\label{sec:Citations_communities_infomap}
\begin{figure}[ht!]
    \centering
    \includegraphics[width=0.98\textwidth]{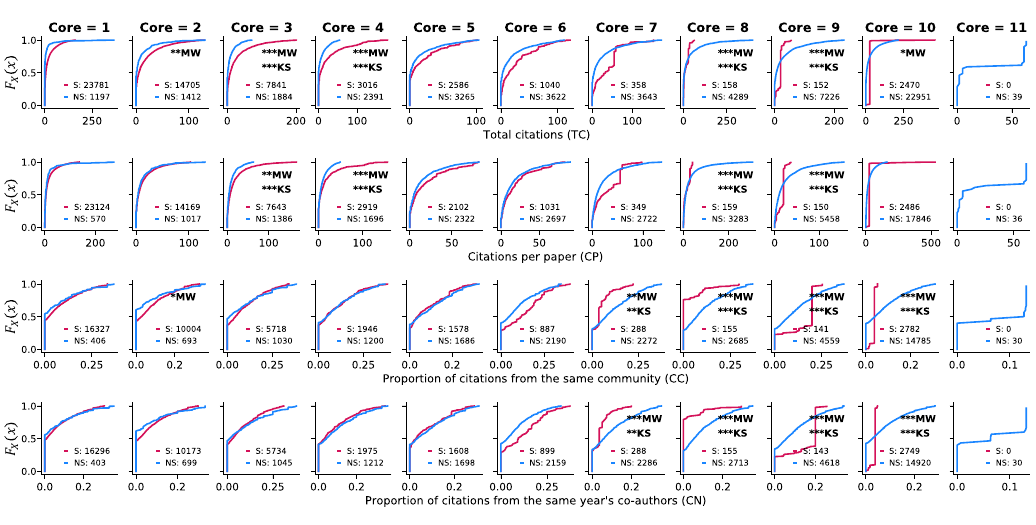}
    \caption{\textbf{Citation metrics for researchers in communities of different segregation categories and core positions for \textit{Infomap} in 2010.} Each row represents the cumulative density function (CDF) for the total citations (TC), the citations per paper (CP), the proportion of citations from the same community (CC), and the proportion of citations from the same year's co-authors (CN). The code of colours is: light red for highly segregated (S) and blue for non-segregated communities (NS). Letters \textbf{KS} or \textbf{MW} appear when there are significant $p$-values values for Kolmogorov-Smirnov (different distribution shapes) and Mann-Whitney (different distribution medians) for the CDFs of non-segregated and highly segregated communities. Significance levels are denoted as follows: * $<$ 0.1, ** $<$ 0.05, and *** $<$ 0.01.}
    \label{fig:Citations_plot_allcores_2010_Infomap}
\end{figure}

 For our main analyses, \textit{Label-propagation} in 2010, we compare the number of papers published with the number of citations and citations per paper received until 2020. In this section, we analyse the first two rows of~\figurename~\ref{fig:Citations_plot_allcores_2010_Infomap}. We found similar results for both algorithms in total citations (TC): highly segregated researchers have more citations in peripheral and middle cores, while non-segregated researchers have more citations in cores near the nucleus. However, for citations per paper (CP), the results differ. In the case of \textit{Infomap}, the highly segregated researchers have more CP in peripheral and middle cores. In contrast, for \textit{Label-propagation}, there are no differences in the CP between non-segregated and highly segregated researchers. And in the nucleus, the non-segregated researchers have higher CP for both algorithms.

Finally, we found that for the proportion of citations from the same community (CC) and the proportion of citations from the same year's co-authors (CN), the results of both algorithms are similar in the nucleus, with non-segregated researchers having higher values of CC and CN than highly segregated researchers. However, the results differ in middle cores. For \textit{Infomap}, non-segregated researchers have higher CC and CN, but for \textit{Label-propagation}, highly segregated researchers have higher values.

Despite differences in middle cores between the algorithms' results, the main argument of this work is not affected: the scientific impact of researchers is consistently influenced by their communities' segregation category and the core position in the nucleus and periphery of the network. 

\newpage
\newpage
\bibliographystyle{bmc-mathphys} 